\documentclass[useAMS,usenatbib]{mn2e}
\usepackage{epsfig}

\def\gtrsim{\ga}

\title[]{
The Morphology of Collisionless Galactic Rings Exterior to Evolving Bars 
}
\author[]{
Micaela Bagley, 
Ivan Minchev,  \&
Alice C. Quillen  \\
Department of Physics and Astronomy, University of Rochester, Rochester, NY 14627} 

\begin{document}
\label{firstpage}
\maketitle

\begin{abstract}
The morphology of the outer rings of early-type spiral galaxies 
is compared to integrations of massless collisionless 
particles initially in nearly 
circular orbits.  Particles are perturbed by a quadrupolar 
gravitational potential corresponding to a growing 
and secularly evolving bar. 
We find that outer rings with R1R2 morphology and pseudorings 
are exhibited by the simulations 
even though they lack gaseous dissipation.  
Simulations with stronger 
bars form pseudorings earlier and more quickly than those with weaker bars.  
We find that the R1 ring, perpendicular to the bar,
is fragile and dissolves after a few bar rotation
periods if the bar pattern speed 
increases by more than $\sim 8\%$, bar strength increases (by $\gtrsim 140\%$) after bar growth, or the bar is too strong ($Q_T>0.3$).  
If the bar slows down after formation, pseudoring morphology
persists and the R2 ring perpendicular to the bar
is populated due to resonance capture. The R2 ring remains
misaligned with the bar and increases in ellipticity as 
the bar slows down.
The R2 ring becomes scalloped and
does not resemble any ringed galaxies
if the bar slows down more than
3.5\% suggesting that bars decrease
in strength before they slow down this much.
We compare the morphology of our simulations to B-band images 
of 9 ringed galaxies from the 
Ohio State University Bright Spiral Galaxy Survey, 
and we find a reasonable match in morphologies to R1R2' pseudorings
seen within a few bar rotation periods of bar formation.
Some of the features previously interpreted 
in terms of dissipative models may be due to transient structure 
associated with recent bar growth and evolution. 
\end{abstract}

\section{Introduction}
Rings in barred galaxies can exist interior to the bar, 
encircling the bar or exterior to the bar. For a review 
on classification and properties of ringed galaxies see \citet{buta96}.
The outer rings of barred galaxies are classified as R1 or R2 depending
upon whether the ring is oriented with major axis
perpendicular to the bar (R1) or parallel
to it (R2) (e.g., \citealt{romero06}).  
If the ring is broken, partial
or is a tightly wrapped spiral it is called a pseudoring 
and denoted R1' or R2'.  Some galaxies contain both types of rings
and are denoted R1R2' or R1R2.
R1' and R2' morphologies were predicted 
as morphological patterns that would be expected near the outer Lindblad
resonance (OLR) with the bar \citep{schwarz81,schwarz84}.
Rings are often the site of active star formation and so 
are prominent in blue visible band images, H$\alpha$ narrow band images,
and HI emission \citep{buta96}.

Orbital resonances, denoted Lindblad Resonances, 
occur at locations in the disk where 
\begin{equation} \Omega_b = \Omega \pm \kappa/m 
\end{equation} 
where $\Omega_b$ is the angular rotation rate of the bar
pattern and $m$ is an integer. 
Here $\Omega(r)$ is the angular rotation rate of a star
in a circular orbit at radius $r$ and $\kappa(r)$ is the 
epicylic frequency.  The $m=2$ OLR is that with $\Omega_b = \Omega + \kappa/2$.
Orbits of stars are often classified in terms of nearby periodic
orbits that are closed in the frame rotating with the bar.
Near resonances orbits become more elongated and have higher epicyclic
amplitudes.
Exterior to the OLR periodic orbits parallel to the bar
are present whereas interior to the OLR both perpendicular and
parallel periodic orbits are present.
For a steady pattern, closed orbits interior to the OLR
are expected to be 
aligned with major axis perpendicular to the bar whereas those exterior
to the OLR are aligned parallel to it (e.g., \citealt{cont89,kalnajs91}).

A common assumption is that rings form because gas accumulates  
at resonances.  This follows as gas clouds cannot follow 
self-intersecting orbits
without colliding.  Because of dissipation in the
gas, the bar can exert a net torque on the gas
leading to a transfer of angular momentum.  The torque is expected
to change sign at resonances so gas can move away from 
them or accumulate at them.
The CR region is expected to be depopulated leading to gas
concentrations at the OLR and ILR resonances.
Gaseous rings form when gas collects 
into the largest periodic orbit near a resonance
that does not cross another periodic orbit \citep{schwarz84}.

\citet{schwarz81,schwarz84} first demonstrated the efficiency of this process. 
Other papers have confirmed and extended this work 
(e.g., \citealt{combes85,byrd94,salo99,rau00,rau04}).
Because dissipation is thought to be important, spiral like features 
and ovals that are not perfectly aligned with
the bar, similar to those observed, are predicted.
In some cases galaxy morphology and kinematics have not been 
successfully modeled with a single steady state bar component.
Improvements in the models have been made with
the addition of
an additional exterior oval or spiral component 
(e.g., \citealt{hunter88,lindblad96}).

Previous work accounting for ring galaxy morphology
has primarily simulated the gas dynamics using sticky particle
simulations that incorporate dissipative or inelastic collisions.   
\citet{rau00} ran N-body stellar simulations  coupled with
sticky gas particles.
These simulations have self-consistent bars so that
the orbits of the stars in the bars are consistent
with the bar's gravitational potential.
The disadvantage of using N-body simulations is that the properties
of the bar such as its pattern speed and strength cannot be set.  
They can only be changed indirectly by varying the initial conditions
of the simulations.
An alternative approach is to set the bar perturbation
strength, shape and pattern speed
and search for likely bar parameters consistent
with the properties of observed galaxies (e.g., \citealt{salo99,rau04,rau08}).

Previous work has explored the affect of bar strength
and pattern speed on ring morphology (e.g., \citealt{salo99,rau04,rau08})
and length of time since the bar grew (e.g., \citealt{rau00,ann00}).
Here we explore the role of bar evolution on
ring galaxy morphology.  By bar evolution we mean
changes in bar pattern speed and strength during and after bar growth.
N-body simulations lacking live halos predict long lived
bars with nearly constant pattern speeds (e.g., \citealt{voglis07}).   
However angular momentum transfer between a bar and the gas disk 
either interior or exterior to the bar or between a bar
and a live halo can cause the pattern speed to
vary (e.g., 
\citealt{debattista98,bournaud02,das03,ath03,sellwood06,martinez06}).
Thus constraints on the secular
evolution of bars could tell us about
the coupling between bars, gas and dark halos.

Gas and stars exterior to a bar are sufficiently distant and moving
sufficiently slowly compared to the bar that
they are unlikely to cause strong perturbations
on the orbits of stars in the bar.
Because a calculation of the gravitational 
potential involves a convolution with an inverse square
law function, high order Fourier components 
are felt only extremely weakly exterior to
the bar. 
The dominant potential term exterior to the bar is the quadrupolar
term which decreases with radius to the third power, $\Phi \propto r^{-3}$.

Here we explore the role of a changing
quadrupolar potential field on the morphology of
stars exterior to a bar.  
In this work we focus on collisionless stellar orbits and 
leave investigating the study of dissipative effects
for future study.  In Section 2 we describe our simulations and 
present the results obtained by varying the parameters.  In Section 3 
we compare the results of our simulations with 9 galaxies from the 
Ohio State University Bright Spiral Galaxy Survey 
(\citet{eskridge02}, hereafter OSUBSGS).  
Finally in Section 4 we summarize and discuss our results.

\section{Test particle simulations}

We perform 2D test-particle simulations of an initially axisymmetric
galactic disk that is perturbed by a forcing bar pattern.  
The rotation curve adopted for a particle in a circular orbit is 
\begin{equation}
v_c(r) =  s^{\gamma/2}
\end{equation}
with $s = \sqrt{r^2 + a^2}$ and $a>0$ a core radius to prevent
extreme orbits near the galaxy center.
A flat rotation curve has $\gamma=0$.
This curve corresponds to an axisymmetric potential 
\begin{equation}
\Phi_0(s) = \left\{
\begin{array}{ll} 
    \log(s)                 & {\rm for} ~~ \gamma=0 \\
    \gamma^{-1} s^{\gamma}  & {\rm for} ~~ \gamma \ne 0
\end{array}
    \right.
\end{equation}

To this axisymmetric component we add
a quadrupole perturbation for the bar
in the form used by \citet{dehnen00,minchev07},
\begin{equation}
\Phi_b(r,\phi,t) = \epsilon \cos\left[ 2 (\phi - \Omega_b t)\right]
\times 
    \left\{ 
\begin{array}{ll} 
           (r_b/r)^3,      & r>r_b,     \\
           2 - (r/r_b)^3,  & r \le r_b. 
\end{array}
\right.
\label{eqn:phib}
\end{equation}
where $r_b$ is the bar length and $\Omega_b$ its angular
rotation rate or pattern speed.

As we wish to explore bars with changing
pattern speeds we allow $\Omega_b$ to vary with time; however,
we fix the ratio of the bar length to the corotation radius, $R$, so
that
\begin{equation}
 r_b(t)  = r_{b,0} { \Omega_{b,0} \over  \Omega_b(t) }.
\end{equation}
Previous studies have found that bars end 
interior to their corotation radius, $r_{CR}$, with
the ratio of bar length  to bar corotation radius 
$R=0.7-0.9$ \citep{ath92,rau08}.   
We describe pattern speed variations with two parameters:
the rate of change during bar growth, $d\Omega_g/dt$, and
that after bar growth, $d\Omega_b/dt$.
The bar strength grows linearly with time, $\epsilon \propto t$,
until a time $t_{grow}$, at which time it reaches a strength 
$\epsilon_{tgrow}$.  
After $t_{grow}$ the bar strength
may vary at a slower rate, $d\epsilon/dt$.

Previous work has used as a measure of bar strength the parameter 
$Q_T$ \citep{combes81}. At a given radius
this is the ratio of
the maximum tangential force to the azimuthally averaged
radial force.
Here equation (\ref{eqn:phib}) implies that
the maximum value of $Q_T$ is $Q_T = 2 \epsilon / v_c^2$.

The simulations presented here integrate $10^5$ particles with a
4$^{th}$ order Runge-Kutta method. 
All particles are integrated simultaneously in parallel
on a NVIDIA GeForce 8800 GTX graphics card.
The code is written with NVIDIA's CUDA (Compute Unified Device Architecture),
the C-language development environment
for CUDA enabled Graphics Processing Units (GPUs).

Particle initial conditions are nearly circular orbits
with epicyclic amplitude randomly generated
so the initial velocity dispersion  is $\sigma$ times the circular
velocity.  In our simulations the velocity dispersion is about 0.04 of the circular velocity, 
which is about 7 km/s for a galaxy with a 200 km/s rotational velocity. 
For comparison, HI line widths are typically  in the range of 5-10 km/s.  
The epicylic amplitude distribution is Gaussian.
Initial radii are chosen from a flat distribution
with minimum and maximum radius between 0.5 and 4.0 times the
initial bar length. This leads to an initial disk surface density
proportional to $1/r$.

Unless otherwise noted, when we discuss times in terms of bar periods 
we are referring to the initial bar period, which has time 
$P_{b,0}=2\pi/\Omega_{b,0}$.  
We run our simulations for twenty-five bar periods, 
and the bar grows for the first three bar rotation periods; $t_{grow}=3$.
We focus on a ratio of bar length to corotation radius of $R=0.8$, 
so that the initial bar pattern speed is $\Omega_{b,0}=0.8$ and the 
corotation radius is therefore $r_{CR}=1.25$.  Table \ref{tab:tab1} 
lists the initial conditions that all of our simulations have in common.  
Table \ref{tab:tab2} lists remaining simulation parameters.
The majority of our simulations have bars of strength 
$|\epsilon_{tgrow}|=0.10$, corresponding to $Q_T = 0.2$.  
Lengths are given in terms of the initial bar length,  
$r_{b,0}$, and angular velocities are given in terms of that at a radius 
of the initial bar length.  We use negative values 
of $\epsilon$ to signify that the bar is initially oriented horizontally.

\subsection{Description of simulations}

\begin{figure}
\includegraphics[angle=0,width=3.3in]{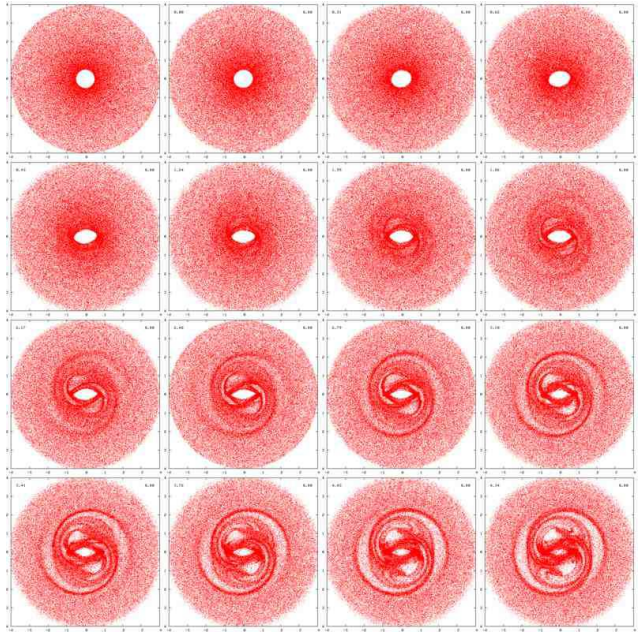}
\caption{
Bar growth in simulation 1.  The distribution of collisionless particles is 
shown every quarter bar period.  The bar is growing up to the left frame 
of the last row.  Strong open spiral-like structure is evident during bar 
growth even though the simulation is lacking gaseous dissipation.
\label{mongrow.eps}
}
\end{figure}

\begin{figure}
\includegraphics[angle=0,width=3.3in]{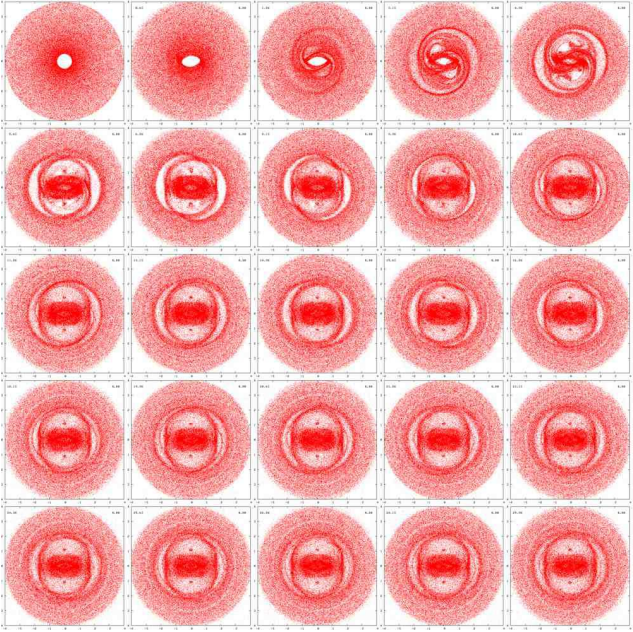}
\caption{
Simulation 1, with bar angular rotation rate
$\Omega_{b,0}=0.8$, bar strength $|\epsilon_{tgrow}|=0.10$  (corresponding
to $Q_T = 0.2$) and bar grown in $t_{grow}=3$ bar periods.
The distribution of collisionless particles is shown each 
full bar rotation period, and the entire 25 bar periods 
of the simulation are shown.
The bar is growing up to the fourth frame.
We note that the R2 ring is misaligned with bar and
azimuthal variations in densities are seen
until $\sim t=5$ bar rotation periods after bar formation.
Pseudoring morphology is present at the end of
and a few rotation periods after bar growth.
Both R1 and R2 rings are present and stable after bar growth.
We find that collisionless particles that are initially
in nearly circular orbits can display R1R2 outer ring morphology.
\label{mon2beven.eps}  
}
\end{figure}

Snapshots at different times for simulation 1
with parameters listed in Tables \ref{tab:tab1} and \ref{tab:tab2}
are shown in Figures \ref{mongrow.eps} and 
\ref{mon2beven.eps}.
Simulation 1 is our base or comparison simulation, 
with $|\epsilon_{tgrow}|=0.10$, corresponding to $Q_T=0.2$,
and bar grown in $t_{grow}=3$ bar rotation periods.  
In Figures \ref{mongrow.eps} and \ref{mon2beven.eps}
each frame has been rotated so that the bar is horizontal.  
Figure \ref{mongrow.eps} shows the first 3 bar 
periods of bar growth of simulation 1 with each image
separated in time by a quarter bar period.  
Figure \ref{mon2beven.eps} shows snapshots of the 25 bar periods 
of the simulation with each image separated by a full bar period.  
As can be seen from Figure \ref{mongrow.eps},
during bar growth, strong open spiral-like structure 
is present that might be interpreted as an R1' ring.
Just after bar growth (see Figure \ref{mon2beven.eps}), 
both R1 and R2 rings are present but the R2 ring is 
not always oriented parallel to the bar.   
For up to 5 bar rotation periods following
bar growth, there are azimuthal variations in density in the rings 
as well as shifts in the R2 ring orientation so they could be 
considered pseudorings.
After bar growth the structure 
stabilizes and R1 and R2 rings remain that are increasingly
mirror symmetric and remain oriented perpendicular and parallel
to the bar, respectively. 

The bar is grown sufficiently slowly that
the orbits change adiabatically.  Orbits remain near to closed 
or periodic orbits and structure
associated with both R1 and R2 orbit families is seen.
Most interesting is that the simulation displays
twists in the density peaks, azimuthal variations in the density
of the ring and deviations of ring orientation from perpendicular and
parallel to the bar at the end of and a few periods after bar growth.  
Previous work has suggested that weak dissipation is required to
exhibit spiral structure or pseudoring morphology, 
however here we see transient spiral 
structures induced by bar growth and pseudoring type morphology
for a few rotation periods following bar growth.
After $\sim 5$ periods the asymmetries are reduced
and the morphology contains both stable R1 and R2 type rings. 

Our simulation looks similar to the sticky particle
simulations by \citet{schwarz81,byrd94}.  Their simulations also
displayed early spiral structure.  The sticky
particle simulations exhibit strong R1' morphology for a few bar
rotation periods.  Our simulation exhibits R1' type morphology
only during bar growth, R1R2'  morphology a few rotation periods
after bar growth and stable R1R2 morphology on long timescales.

During bar growth these simulations look remarkably similar
to the sticky particle simulations by \citet{rau04,rau08} even
though they lack dissipation.
N-body and SPH simulations often suffer from 
artificially heating. Fine structure that would
only be present in a perturbed initially cold population might
not survive due to excess heating inherent in the simulation.
Sticky particle and SPH simulations, because they allow dissipation,
can reduce the velocity dispersion of the particles. 
The similarity between the sticky particle simulations
and the dissipationless simulations shown here could be 
because our initial orbits were nearly circular.

The ability of collisionless simulations to display R1R2 type morphology
suggests that we reexamine the role of dissipation
in influencing ring galaxy morphology.
SPH simulations \citep{ann00,bissantz03} sometimes show open outer
spiral arms that are similar to R1' pseudorings and resemble
morphology during bar growth seen here.  
However our simulations stop showing spiral structure 
soon after bar growth.
The SPH simulations by \citet{ann00} show R1' type pseudoring morphology
but only within a few rotation periods after bar growth.
Likewise the sticky particle simulations by \citet{rau00} show R1 or R2
or both R1R2
ring morphology, but only within a few bar rotation periods after bar growth
(see their figure 10).
The SPH simulations by \citet{ann00}
show spiral structure for a somewhat longer time
than ours (a few bar rotations following bar growth) 
but they fail to exhibit R1 or R2 ring morphology.

We find that R1R2 rings (systems with both types of features) 
do not require dissipation for formation, however they do require particles
to be on nearly closed orbits.  This can result either
because of dissipation or because gas and recently born stars
tend to be on nearly circular orbits prior to bar growth.
The success of sticky particle simulations in 
reproducing outer ring morphology, may be in part because of
their ability to cool or reduce the velocity dispersion of their particles.
SPH simulations (e.g., \citealt{ann00,bissantz03}) exhibit
only R1' pseudoring morphology suggesting that when dissipation
is large, both types of rings are not formed.
Recent SPH and sticky particle simulations fail to exhibit
long lived R1R2 morphology.  Here however
we see that R1R2 morphology can be long lived, though 
it's possible that R1R2 ring morphology 
is a short lived phenomenon as many barred galaxies do not exhibit
R1 or R2 type outer rings.

\subsection{Morphology sensitivity to pattern speed variation after
bar growth}

\begin{figure}
\includegraphics[angle=0,width=3.3in]{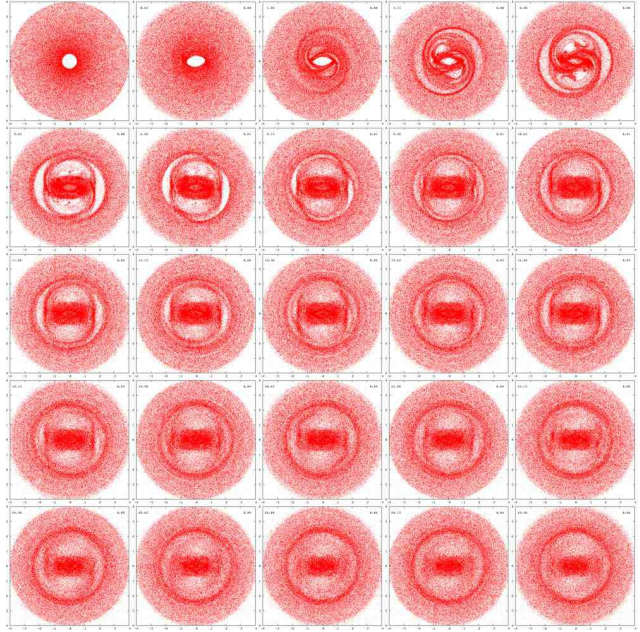}
\caption
{Simulation 5 showing a bar with increasing pattern
speed, $d\Omega_b/dt=0.0004$.  The bar speeds up only after 
it has finished growing at 3 bar rotation periods.  
Each frame is separated by one full initial bar 
rotation period.  Note the loss of the R1 ring later in the simulation.
\label{mon9beven.eps} 
}
\end{figure}

We first explore the sensitivity of outer ring morphology to
weak changes in bar pattern speed following bar growth. 
Simulations 2-6 are identical to simulation 1 except the bar
pattern speed increases after bar growth.  Simulations 7-11 are 
identical to simulation 1 except the bar pattern speed decreases after
bar growth.
Figure \ref{mon9beven.eps} shows simulation 5 that has 
an increasing bar pattern speed $d\Omega/dt = 0.0004$.  
For this simulation the pattern speed increases 0.39\% each bar period.
We find that the R1 outer 
ring grows weaker as the pattern speed increases.  The R1 
has completely dissolved   by the end of the simulation when the pattern
speed has increased by about 9\% compared to its initial value.
At later times, even though the perturbation is always changing the morphology
is nearly  mirror symmetric.  We find that spiral structure or pseudoring
structure is not caused by the increase in bar pattern speed.

\begin{figure}
\includegraphics[angle=0,width=3.3in]{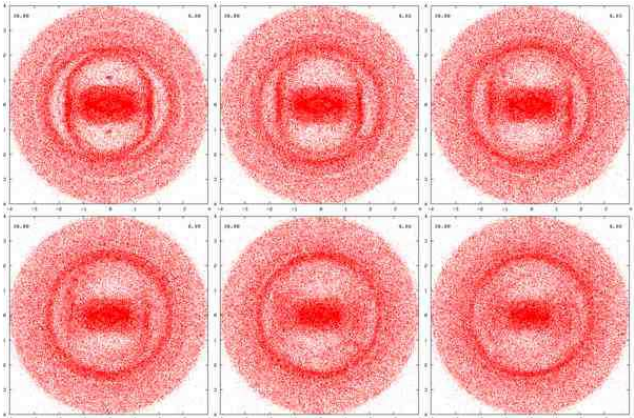}
\caption{
The last frame of simulations 1-6.  Each frame shows the density distribution
at a time 25 bar periods after the start of the simulation.
From left to right the 
top row shows simulation 1 with $d\Omega_b/dt=0.0$; simulation 2 with
$d\Omega_b/dt=0.0001$ and simulation 3 with $d\Omega_b/dt=0.0002$.  
The bottom row shows simulation 4 with $d\Omega_b/dt=0.0003$; 
simulation 5 with $d\Omega_b/dt=0.0004$ and simulation 6 with $d\Omega_b/dt=0.0005$.  
We find that when the  bar pattern speed increases by 
more than $\sim 8\%$ the R1 ring completely dissolves.
\label{domegab.eps} 
}
\end{figure}

Figure \ref{domegab.eps} shows the morphology
at the end of simulations 1-6.  
Here we see that the R1 ring disappears as the bar pattern
speed increases. 
At later times in the simulations with more quickly
increasing pattern speeds the R1 ring dissolves and only a nearly circular
outer ring remains.
We find that when the  bar pattern speed increases by 
more than $\sim 8\%$, the R1 ring completely dissolves.

\begin{figure}
\includegraphics[angle=0,width=3.3in]{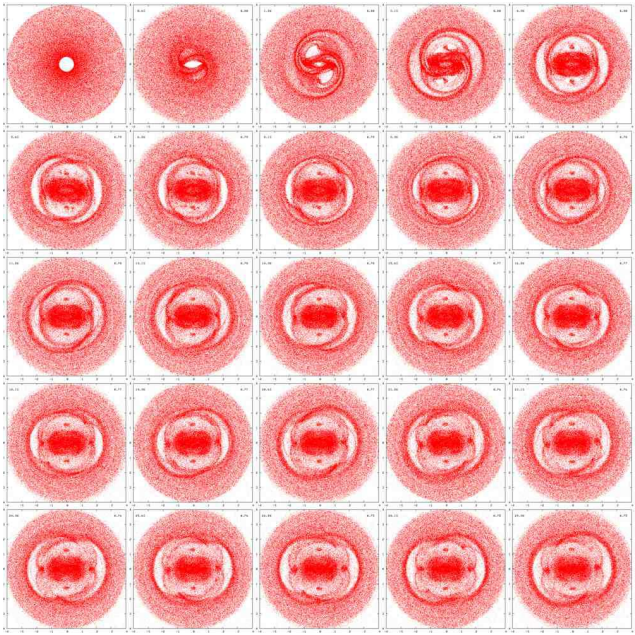}
\caption{
Simulation 11 showing a bar with decreasing pattern speed,
$d\Omega_b/dt=-0.0003$.  The bar slows down
after it has finished growing.  Each frame is separated by one
full bar rotation period.  We see a high epicyclic amplitude
R2 ring that we attribute to resonance capture.
This ring can be misaligned with the bar
even at late times.
\label{mon14beven.eps}
}
\end{figure}

In simulations 7-11 we decrease the bar pattern speed after bar growth.  
Figure \ref{mon14beven.eps} shows simulation 11 with
$d\Omega_b/dt=-0.0003$.
The R2 ring in this simulation is elongated and strong and is seldom aligned
parallel to the bar even at later times.  While the simulation
with the increasing pattern speed (shown in Figure \ref{mon9beven.eps})
did not exhibit pseudoring morphology after bar growth
or misaligned R1 or R2 rings, the simulations with decreasing
pattern speed do show misaligned R2 type rings even at late times. 
The ellipticity of the R2 ring is higher than seen in the comparison
simulation shown in Figure \ref{mon2beven.eps} with a bar with a constant
pattern speed.

\begin{figure}
\includegraphics[angle=0,width=3.3in]{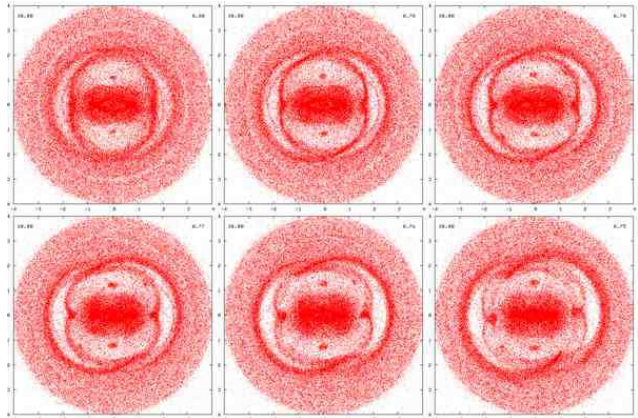}
\caption{
The last frame of simulations 1 and 7-11.  Each frame shows the 
density distribution at a time 25 bar periods after the start 
of the simulation.  From left to right the top row shows 
simulation 1 with $d\Omega_b/dt=0.0$, simulation 7 with 
$d\Omega_b/dt=-0.0001$ and simulation 8 with $d\Omega_b/dt=-0.00015$.  
The bottom row shows simulation 9 with $d\Omega_b/dt=-0.0002$, 
simulation 10 with $d\Omega_b/dt=-0.00025$ and simulation 11 with 
$d\Omega_b/dt=-0.0003$.  We attribute the high epicyclic amplitude 
R2 rings to resonance capture.  The R2 ring remains misaligned with 
the bar even 22 bar periods after the bar pattern speed began decreasing.  
\label{domegab2.eps} 
}
\end{figure}

Figure \ref{domegab2.eps} shows the last periods of 
simulations 1 and 7-11.  Here we see that the R2 ring 
remains misaligned 22 bar periods after bar growth is 
complete when the pattern speed begins to decrease.  
The ellipticity of the R2 ring increases as the bar pattern 
speed decreases.  We find that when the pattern speed 
decreases by more than $\sim 3.5\%$, the simulations 
do not resemble real galaxy morphology.  

Decreasing the pattern speed moves the bar's resonances  outwards.  
Exterior to the OLR only one family of periodic orbits exits
aligned parallel to the bar.  However interior to the bar,
two families of periodic orbits exist,
both those perpendicular and parallel to the bar 
(e.g., \citealt{cont89,quillen03}).
When the bar slows down orbits can be captured into resonance
(e.g., \citealt{romero06}).
Only the orbit family parallel to the bar can capture particles,
and as the bar pattern speed continues to decrease these orbits
will increase in epicyclic amplitude.
This is mathematically similar to Pluto's orbit increasing 
in eccentricity as Neptune migrates outwards (e.g., \citealt{quillen06}).
Resonance capture into orbits parallel with the bar is likely to
explain the higher epicyclic amplitude of the R2 rings seen
in Figure \ref{mon14beven.eps} in which the bar pattern
is slowing down.

In summary if the bar speeds up subsequent to bar formation 
(or increases in pattern speed by more than $\sim 8\%$)
we find that the R1 ring dissolves.  
Simulations with increasing pattern speeds show
misaligned rings and azimuthal density contrasts
only 1-2 periods after bar formation.
If the bar slows down however, we find that the R2 ring
is increased in strength and eccentricity and both R2 and R1 are
seen even at later times ($>20$ periods after  bar formation).
Furthermore the R2 ring is misaligned with the bar for
many rotation periods.  If the bar decreases by more than
$\sim 3.5\%$ we find that the R2 becomes unrealistically scalloped. 

\subsection{Morphology sensitivity to pattern speed variation during
bar growth}

\begin{figure}
\includegraphics[angle=0,width=3.3in]{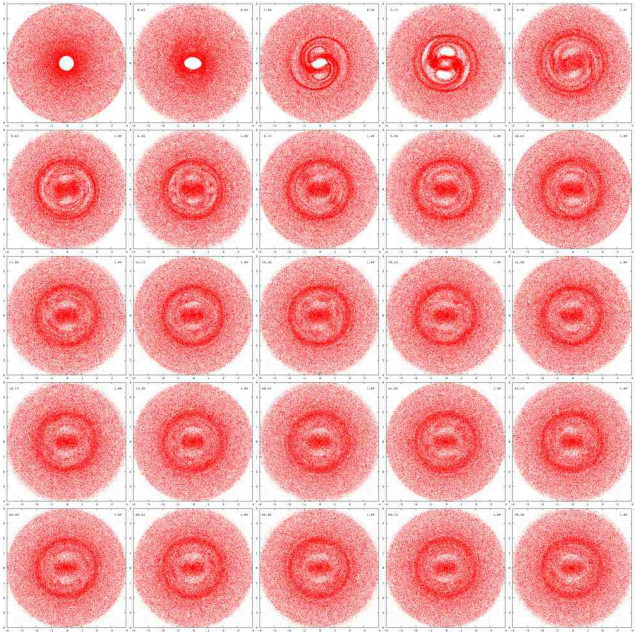}
\caption{
Simulation 13 showing a bar with increasing pattern speed during bar growth, 
$d\Omega_g/dt=0.01$.  The bar pattern speed is only increasing in the 
first three frames while the bar is growing.  Each frame is separated 
by one full initial bar period.  R1R2 structure does not form as it 
does in Figure \ref{mon9beven.eps} where the pattern speed is increasing 
after bar growth.  Here spiral structure dissolves within 3 periods 
after bar growth and only a nearly circular outer ring remains.
Outer ring structures may be more sensitive to alterations 
in pattern speed during bar growth than to changes after bar growth.
\label{mon21beven.eps} 
}
\end{figure}

We next explore the effect of altering the bar pattern 
speed during bar growth rather than following bar growth.  
$d\Omega_g/dt$ is positive in simulations 12-15, causing the 
pattern speed to increase for the first 3 periods of each simulation.  
Figure \ref{mon21beven.eps} shows simulation 13 with $d\Omega_g/dt=-0.01$.  
Open spiral arms are present at the end of the third period of bar growth.  
However, R1R2 double ring morphology does not form in
this simulation. 
As was true in the simulations with bar pattern speed increasing following
bar formation (simulations 2-6; see Figures \ref{mon9beven.eps} and
\ref{domegab.eps}) 
the R1 ring dissolves and only a nearly circular outer ring remains.  
We note that the spiral and outer ring structure in simulations 2-6 
(see \ref{mon9beven.eps}) did not dissolve as quickly as in simulation 13, 
where an R1R2 outer ring never forms.  This suggests that 
outer ring structures are more sensitive to alterations 
in pattern speed during bar growth than to changes after bar growth.

\begin{figure}
\includegraphics[angle=0,width=3.3in]{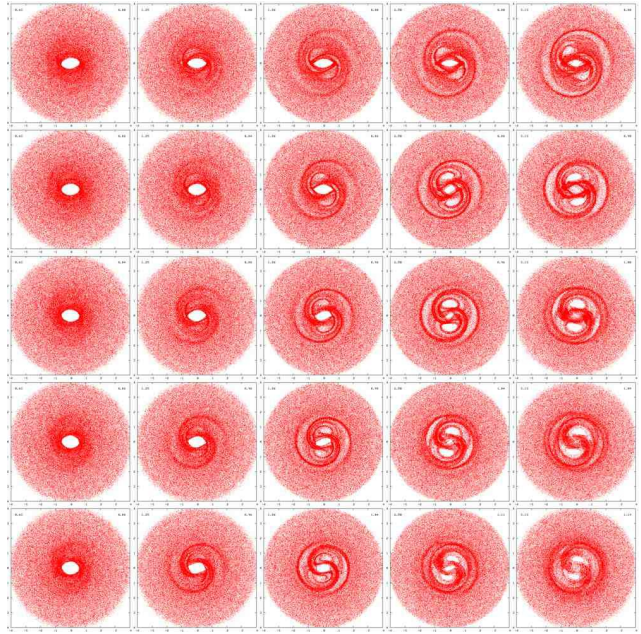}
\caption{
Morphology change when the pattern speed is increased during bar growth.
Five frames, corresponding to the last 2 periods of bar growth, 
of the simulations in which the pattern speed is increased 
during growth.  From top to bottom the rows show simulation 1 with 
$d\Omega_g/dt=0.0$; simulation 12 with $d\Omega_g/dt=0.005$; simulation 13 with 
$d\Omega_g/dt=0.01$; simulation 14 with $d\Omega_g/dt=0.015$ and simulation 15 with 
$d\Omega_g/dt=0.02$.  As the pattern speed increases, the axes of the rings 
decreases.
\label{domdtgr.eps} 
}
\end{figure}

Figure \ref{domdtgr.eps} shows 
the last 2 of the 3 periods of bar growth for simulations 1 and 12-15.  
The first row shows simulation 1 with $d\Omega_g/dt=0.0$.  $d\Omega_g/dt$ 
increases by 0.005 in each consecutive simulation, corresponding
to each row in Figure \ref{domdtgr.eps}. 
We note from these
simulations that open spiral arms are seen during bar growth when the bar is
increasing in pattern speed.  
The radii of spiral structure decreases as the
pattern speed increases.  This is expected as
the radii of the 
resonances move inward as the bar pattern speed increases. 

Figure \ref{domegab.eps} shows 
that ring radial size does not decrease as
the bar pattern speed increases
when the pattern speed increases after bar growth.
However when the pattern speed increases during bar
growth (see Figure \ref{domdtgr.eps}) the ring radius 
does decrease.
This suggests that ring size is primarily set during bar growth
and is not strongly affected by subsequent increases in pattern speed.
Subsequent bar speed increases primarily dissolves or weakens 
the rings rather than changes their radius.

\begin{figure}
\includegraphics[angle=0,width=3.3in]{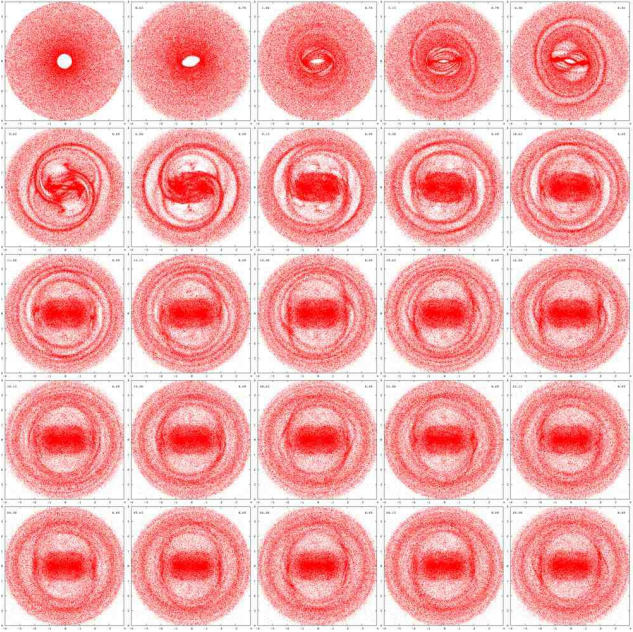}
\caption{
Simulation 17 with $d\Omega_g/dt=-0.005$.  
The pattern speed is decreasing during bar growth.  
Each frame is separated by one full bar rotation period.  
The transient spiral structure during bar growth 
is weaker than when bar pattern is fixed 
(see top row of Figure \ref{mon2beven.eps}).
Formation of the R1 and R2 rings is delayed in this simulation 
and occurs after the bar has finished growing compared
to that with a fixed pattern speed.
\label{mon43beven.eps} 
}
\end{figure}

\begin{figure}
\includegraphics[angle=0,width=3.3in]{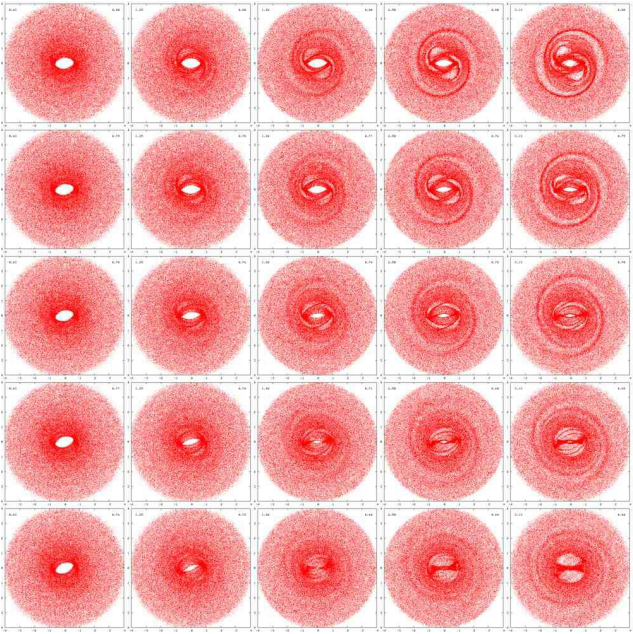}
\caption{
Morphology change when the pattern speed is 
decreased during bar growth.  Five frames, 
corresponding to the last 2 periods of bar growth,
of the simulations in which the pattern speed is 
decreased during bar growth.  From top to bottom the
rows show simulation 1 with $d\Omega_g/dt=0.0$; 
simulation 16 with $d\Omega_g/dt=-0.0025$;
simulation 17 with $d\Omega_g/dt=-0.005$; 
simulation 18 with $d\Omega_g/dt=-0.0075$; and 
simulation 19 with $d\Omega_g/dt=-0.01$.
Transient spiral structure during bar growth is weaker when
the bar is slowing down. 
\label{domdtgr2.eps} 
}
\end{figure}

We now compare the effect of decreasing pattern speed during
bar growth with the effect of increasing pattern speed during bar
growth.
The pattern speed is decreasing during the 3 periods of bar growth 
in simulations 16-19.  Figure \ref{mon43beven.eps} shows the entire simulation 
17 with $d\Omega_g/dt=-0.005$, while Figure \ref{domdtgr2.eps} 
shows the last 2 of the 3 periods of bar growth for simulations 1 and 16-19.  
We find that when the bar pattern slows down during bar
growth the outer rings are not lost as was true when the 
pattern speed increased during growth. 
The R1R2' structure 
seen during bar growth for simulation 1 is exhibited by
this simulation but later, 2 to 3 periods after bar growth 
rather than in the first period following bar growth.  
During this time the rings appear almost double or tightly wound.
The pseudorings close, 
the morphology stabilizes and presents 
the R1R2 structure characteristic of our steady state 
comparison simulation shown in Figure \ref{mon2beven.eps}.  
As is expected from the location
of the OLR, the ring radii become larger if the
bar pattern speed decreases. 

The first row of Figure \ref{domdtgr2.eps} shows simulation 1 with 
$d\Omega_g/dt=0.0$.  $d\Omega_g/dt$ decreases by 0.0025 in each 
consecutive simulation, corresponding to each row in Figure \ref{domdtgr2.eps}. 
The decreasing pattern speed seems to delay the formation of strong spiral
structure. 
Figure \ref{domdtgr.eps} showing morphology during bar growth for a 
bar that is speeding up can be compared to Figure \ref{domdtgr2.eps} 
that shows morphology during bar growth for a bar that is slowing down.  
The simulation with the largest decrease
in pattern speed shows the weakest spiral structure during bar growth
and that with the largest increase in pattern speed the strongest spiral
structure earliest.

In summary, we find that if the bar pattern speed is decreasing during growth,
transient spiral structure is weaker during growth and the formation 
of the R1 and R2 ring structure is delayed by a few bar rotation periods.  
If the bar pattern speed increases during rather than after growth, 
the outer rings are smaller.  An increase in pattern speed during bar growth
destroys the R1 ring and asymmetries typical of pseudorings.

\subsection{Morphology sensitivity to bar strength}

\begin{figure}
\includegraphics[angle=0,width=3.3in]{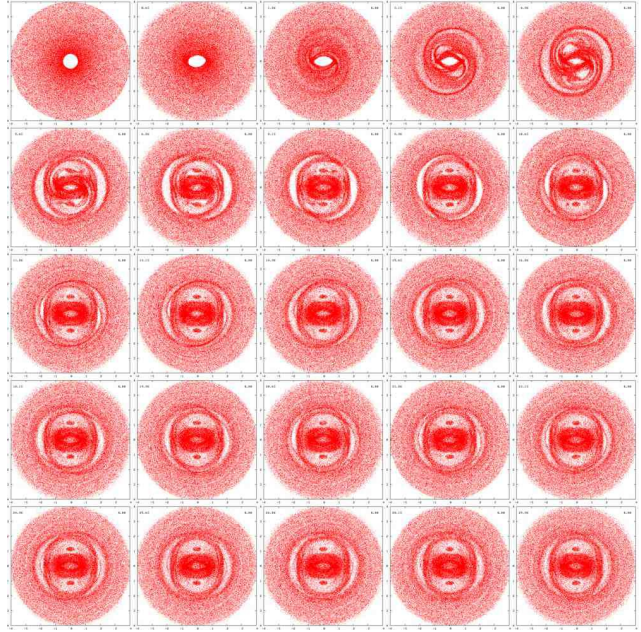}
\caption{
Simulation 21 for a weak bar with $|\epsilon_{tgrow}|=0.08$ 
corresponding to $Q_T=0.16$.  This is a weaker bar than that 
of simulation 1 shown in Figure \ref{mon2beven.eps} 
where $|\epsilon_{tgrow}|=0.10$.  
Each frame is separated by one full bar period.  
The R2 ring is misaligned with the bar for up to 6 bar periods 
after bar growth, longer than the R2 ring in Figure \ref{mon2beven.eps}.  
At late times stable R1 and R2 rings form and remain mirror symmetric.
\label{mon46beven.eps} 
}
\end{figure}

We now explore the sensitivity of the morphology to bar strength.
The bar strength increases linearly with time during bar growth until, 
at $t=t_{grow}$, 
it reaches a strength determined by the parameter $|\epsilon_{tgrow}|$.  
Thus a lower value of $|\epsilon_{tgrow}|$ results in both a weaker 
bar at $t=t_{grow}$ 
and a slower rate of bar growth.  Figure \ref{mon46beven.eps} 
shows simulation 21 with $|\epsilon_{tgrow}|=0.08$ corresponding to $Q_T=0.16$.
This is a weaker bar than that of our comparison simulation shown 
in Figure \ref{mon2beven.eps} where $|\epsilon_{tgrow}|=0.10$.  
During and right after bar growth we see the open spiral structure evident 
in Figure \ref{mon2beven.eps}.  Here the R2 ring is misaligned with the bar 
for up to 6 bar rotation periods following bar growth, whereas the 
R2 ring in Figure \ref{mon2beven.eps} is misaligned for 5 bar periods 
following bar growth.   The timescale for misalignment in the R2 ring 
is probably related to the bar strength, which would set the libration
timescale in the OLR.
The R1 and R2 rings become increasingly mirror 
symmetric.  They remain strong, stable, and oriented perpendicular and parallel 
to the bar, respectively.   

\begin{figure}
\includegraphics[angle=0,width=3.3in]{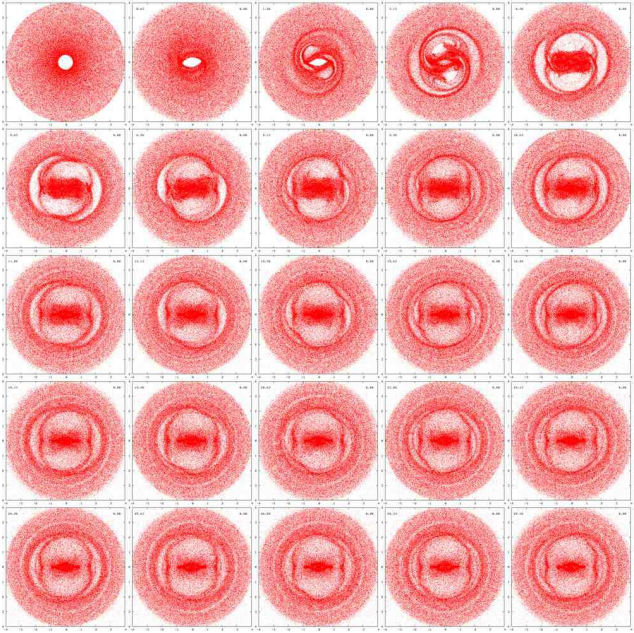}
\caption{
Simulation 23 for a strong bar with $|\epsilon_{tgrow}|=0.14$ corresponding 
to $Q_T=0.28$.  This is a stronger bar than that shown 
in Figure \ref{mon2beven.eps}.  Each frame is separated 
by one full bar period.  The R2 ring is only misaligned with the 
bar for 3 bar periods after bar growth.  The R1R2 structure 
is evident earlier than in Figure \ref{mon46beven.eps} but the 
R1 ring weakens at the end of the simulation.
\label{mon38beven.eps} 
}
\end{figure}

Figure \ref{mon38beven.eps} shows simulation 23 with $|\epsilon_{tgrow}|=0.14$ 
corresponding to $Q_T=0.28$.  This is a stronger bar than that of our 
comparison simulation shown in Figure \ref{mon2beven.eps}.  
The spiral structure evident during and right after bar growth is much 
stronger than that of either Figure \ref{mon46beven.eps} or 
Figure \ref{mon2beven.eps}.  The R2 ring is misaligned with the bar 
for only 3 to 4 bar periods following bar growth. R1 and R2 rings 
form earlier, and we see a weakening of the R1 ring at later
times of the simulation.

\begin{figure}
\includegraphics[angle=0,width=3.3in]{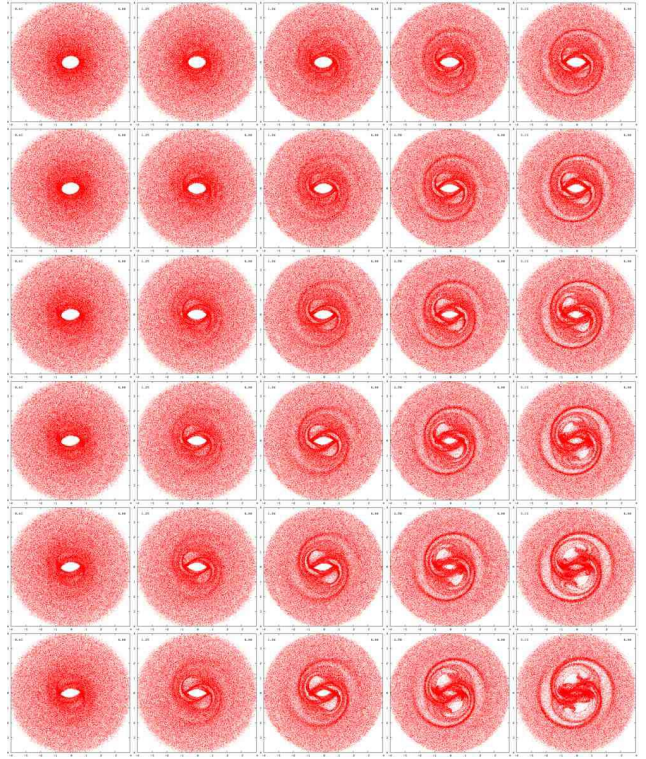}
\caption{
Morphology change when the bar strength is altered.  Five frames, 
showing the last 2 periods of bar growth, of simulations 1 and 20-24.  
From top to bottom the rows show 
simulation 20 with $|\epsilon_{tgrow}|=0.06$, $Q_T=0.12$; 
simulation 21 with $|\epsilon_{tgrow}|=0.08$, $Q_T=0.16$; 
simulation 1 with $|\epsilon_{tgrow}|=0.10$, $Q_T=0.2$; 
simulation 22 with $|\epsilon_{tgrow}|=0.12$, $Q_T=0.24$; 
simulation 23 with $|\epsilon_{tgrow}|=0.14$, $Q_T=0.28$; 
and simulation 24 with $|\epsilon_{tgrow}|=0.16$, $Q_T=0.32$.  
By the end of the fifth frame, 
$t=t_{grow}$ and the strength of the bar is equal to the value of the parameter 
$|\epsilon_{tgrow}|$.  Bars with higher values of $|\epsilon_{tgrow}|$, 
i.e. stronger bars that grow faster, develop strong spiral 
structure and outer rings earlier.  
By the end of the 3 periods of bar growth, our strongest bar 
(simulation 24, 
$|\epsilon_{tgrow}|=0.16$) has pseudorings that have almost fully 
closed to form an R1R2 ring.
\label{eps1.eps} 
}
\end{figure}

In Figure \ref{eps1.eps} 
we compare the last 2 periods of bar growth of simulation 1 
and simulations 20-24.  
$|\epsilon_{tgrow}|$ increases from 0.06 (top row of Figure \ref{eps1.eps}) 
to 0.16 (bottom row).  
Strong transient spiral structure forms noticeably earlier 
when the bar is stronger.

By the third period of simulation 24 in the bottom row, 
$|\epsilon_{tgrow,24}|=0.16$ ($Q_T=0.32$), 
the pseudoring structure most closely resembles that of the fourth 
period of simulation 1, 
$|\epsilon_{tgrow,1}|=0.10$.  Simulation 24 exhibits closed R1R2 
structure a half period 
after bar growth, whereas simulation 1's R1R2 structure is not evident 
until 1.5 periods 
after bar growth.  Thus, a 60\% increase in bar strength accelerates pseudoring 
formation such that closed-orbit rings are evident one bar period earlier.    

\begin{figure}
\includegraphics[angle=0,width=3.3in]{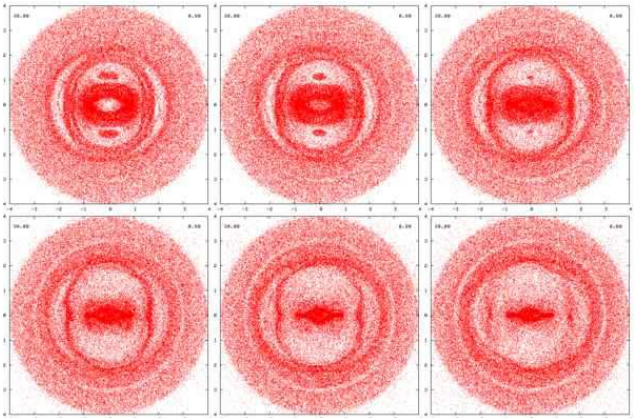}
\caption{
The last frame of the simulations in which the strength
of the bar, $|\epsilon_{tgrow}|$, is varied.  Each frame shows 
the density distribution at a time 25 bar periods after the 
start of the simulation. This figure shows the same simulations 
as Figure \ref{eps1.eps}.  
From right to left the top row shows simulation 20 with 
$|\epsilon_{tgrow}|=0.06$, simulation 21 with $|\epsilon_{tgrow}|=0.08$ 
and simulation 1 with $|\epsilon_{tgrow}|=0.10$.  
The bottom row shows simulation 22 with $|\epsilon_{tgrow}|=0.12$, 
simulation 23 with $|\epsilon_{tgrow}|=0.14$ and simulation 24 with 
$|\epsilon_{tgrow}|=0.16$.  
We find that R1 rings dissolve at later times when the bar strength 
$|\epsilon| > 0.15$.
\label{eps3.eps}
} 
\end{figure}

We now consider the structure at later times as a function 
of bar strength.
Figure \ref{eps3.eps} shows the last bar period of each of the 
same six simulations, 
simulations 1 and 20-24.  The bar strength, $|\epsilon_{tgrow}|$, 
increases by 0.02 
in each successive simulation.  
In the first and second frames we see a very strong concentration 
of particles in the R1 ring.  
By the last frame in the second row, however, the R1 ring is almost 
completely gone.  
Thus, though the increased bar strength accelerates ring formation 
at early times, we find that the R1 outer ring
dissolves at later times when the bar strength is high, just as 
it does in Figure \ref{mon9beven.eps} when the pattern speed is 
increasing after bar growth.

The concentrations of particles above and below the bar that are evident
in Figure \ref{mon46beven.eps} and in the weaker bars shown at later 
times in Figure \ref{eps3.eps}
are the L4 and L5 Lagrange points and so are corotating
with the bar.  Small changes in the orbits circulating
around these points would
cause the particles to circle the bar rather than remain
confined to the vicinity of the L4 or L5 point
corotating with the bar (e.g., \citealt{cont06}).
Small variations in pattern speed during or after bar growth (e.g., 
see Figures \ref{domegab.eps} and \ref{mon21beven.eps}) 
and stronger bars (e.g., Figure \ref{eps3.eps}) reduce the number of particles
near these points.  We note that
stars and gas are not commonly seen in galaxies at these points, 
suggesting that weak bars with unchanging pattern speeds and
strengths do not persist in galaxies.

\subsection{Morphology sensitivity to slow variations 
in bar strength after bar growth}

\begin{figure}
\includegraphics[angle=0,width=3.3in]{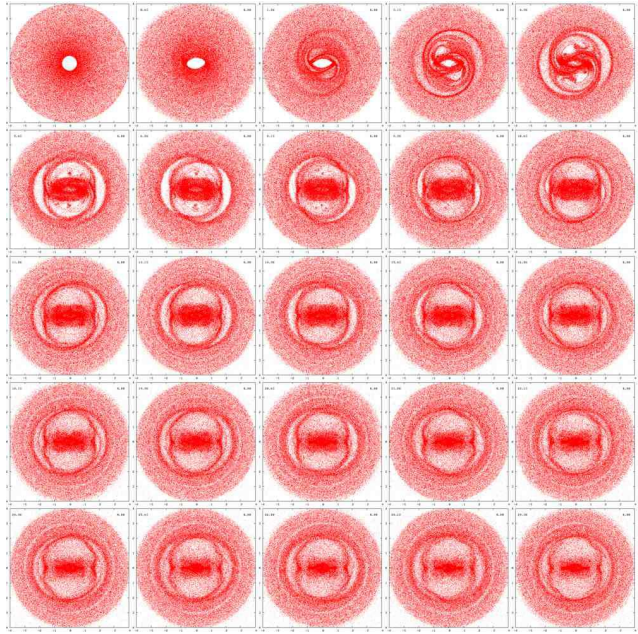}
\caption{
Simulation 26 showing a bar with strength increasing after bar 
growth, $d|\epsilon|/dt= 0.0004$.  Each frame is separated by one 
full initial bar rotation period.  As was true
for the strong bar shown in Figure \ref{mon38beven.eps}, 
increasing bar strength causes the R1 ring to weaken and dissolve.
\label{mon26beven.eps} 
}
\end{figure}

We now consider variations in bar strength after bar growth.
In simulations 25-29, we allow the bar strength to grow linearly 
with time after $t=t_{grow}$.  
Figure \ref{mon26beven.eps} 
shows simulation 26 with increasing bar strength. 
The bar strength at the end of bar growth is $|\epsilon_{tgrow}|=0.10$.  It then continues to increase in 
strength at a slower rate with $d|\epsilon|/dt=0.0004$. 
The bar strength increases by 3.1\% each rotation period reaching
a final value of $|\epsilon| = 0.17$. 
As was true for the strong bar shown in Figure \ref{mon38beven.eps}, 
the R1 ring weakens as the simulation progresses.  

\begin{figure}
\includegraphics[angle=0,width=3.3in]{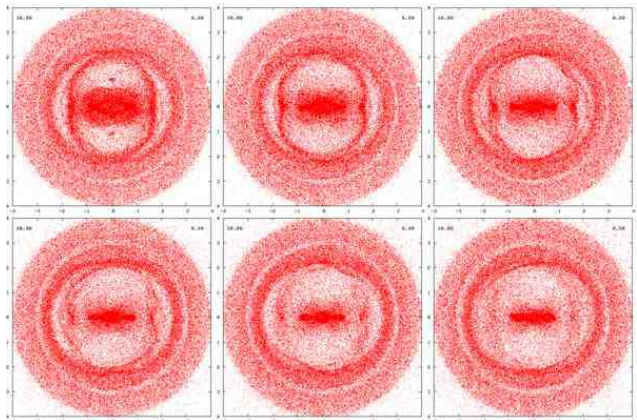}
\caption{
The last frame of each of the simulations in 
bar strength is increasing after $t=t_{grow}$.  
Each frame shows the density distribution at a time 25 bar periods 
after the start of the simulation.  From right to left the top row shows 
simulation 1 with $d|\epsilon|/dt=0.0$, 
simulation 25 with $d|\epsilon|/dt= 0.0002$ and 
simulation 26 with $d|\epsilon|/dt= 0.0004$.  
The bottom row shows 
simulation 27 with $d|\epsilon|/dt= 0.0006$, 
simulation 28 with $d|\epsilon|/dt= 0.0008$ and 
simulation 29 with $d|\epsilon|/dt= 0.0010$.  
We see the same loss of the R1 ring 
as that in Figure \ref{eps3.eps} (showing
the morphology at later times as a function
of bar strength) suggesting
that bars with $|\epsilon| \gtrsim 0.16$ cannot
maintain stable R1 rings. 
\label{depsdt2b.eps}
}
\end{figure}

Figure \ref{depsdt2b.eps} shows the last frame of simulation 1 and 25-29
that have different rates of change in the bar strength following bar
growth.  
We find that by the twenty-fifth bar period, the density distributions 
of these simulations exhibit the same loss of the R1 ring 
as the simulations in which we increased the value of 
the bar strength $|\epsilon_{tgrow}|$.  

We can conclude that an increase in bar 
strength dissolves the R1 ring.  
We lose the R1 ring if the bar becomes 
60\% stronger ($|\epsilon_{tgrow}|=0.16$ 
as opposed to $|\epsilon_{tgrow}|=0.10$) or if the bar strength 
increases by $\sim 140\%$ compared to its initial value.  
It does not appear to matter whether this increase in bar strength occurs 
during bar growth or more slowly after bar growth is completed; 
we find that the R1 ring dissolves in both cases.  
Thus, the R1 ring dissolves for bars with higher strengths.  

\begin{figure}
\includegraphics[angle=0, width=3.3in]{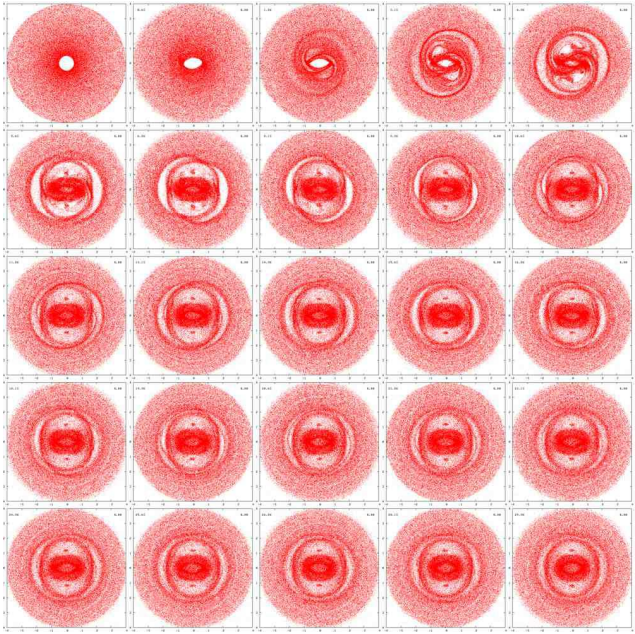}
\caption{
Simulation 32 showing a bar with decreasing strength, 
$d|\epsilon|/dt=-0.0003$.  
The bar strength decreases linearly with time after bar growth is 
complete.  
Each frame is separated by one full initial bar period.  The R1 and R2 
rings become more circular as the bar weakens.  
\label{mon52beven.eps}
}
\end{figure}

\begin{figure}
\includegraphics[angle=0,width=3.3in]{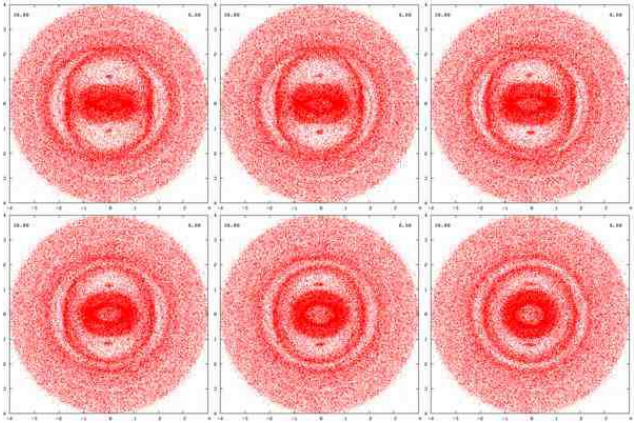}
\caption{
The last frame of each of the simulations in which 
the bar is decreasing in strength after growth. 
Each frame shows the density distribution at a time 25 bar periods 
after the start of the simulation.  From left to right the top row 
shows 
simulation 1 with $d|\epsilon|/dt=0.0$, 
simulation 30 with $d|\epsilon|/dt=-0.0001$ and 
simulation 31 with $d|\epsilon|/dt=-0.0002$.  
The bottom row shows 
simulation 32 with $d|\epsilon|/dt=-0.0003$, 
simulation 33 with $d|\epsilon|/dt=-0.0004$ and 
simulation 34 with $d|\epsilon|/dt=-0.0005$.  
As the bar becomes weaker the rings separate
and become more circular.  
Two circular rings are left at the end of the simulations
that have the weakest bars at the end of the simulation.
NGC 2273 with  an unusual double outer ring \citep{buta96} may be an
example of a galaxy with a bar that has weakened.
\label{depsdt2c.eps} 
}
\end{figure}

We now consider simulations 30-34 in which the bar strength decreases 
after bar growth.  Figure \ref{mon52beven.eps} shows simulation 32 
with $d|\epsilon|/dt=-0.0003$, while Figure \ref{depsdt2c.eps} 
shows the last frame of simulations 1 and 30-34.  

In Figure \ref{mon52beven.eps} showing a bar that
decreases in strength after growth, we see the strong open spiral 
structure during and immediately following bar growth displayed
by many of our simulations.  The R2 ring 
is misaligned with the bar for $\sim 5$ bar periods after bar growth.  
After the R1 and R2 rings form, they begin to lose their respective 
alignments (perpendicular and parallel) with the bar in favor of more 
circular orientations.  As shown in Figure \ref{depsdt2c.eps}, we 
find that the outer rings become circular at later times when the bar 
strength is decreased after bar growth.  
The simulations with the weakest bars at the end of
the simulation leave behind two circular rings.   NGC~2273
an unusual double-outer ring galaxy \citep{buta96},  may be an example
of a bar that has weakened. 

\subsection{Morphology sensitivity to the ratio of bar length 
to corotation radius}

\begin{figure}
\includegraphics[angle=0,width=3.3in]{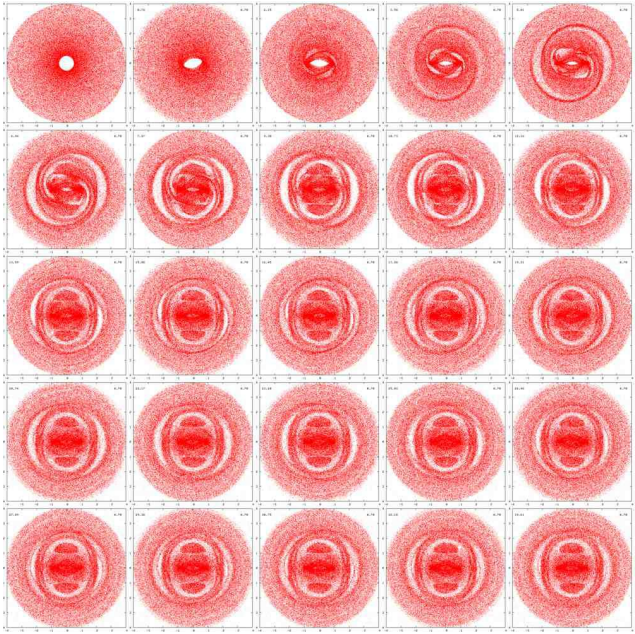}
\caption{
Simulation 35 with a slower initial pattern speed of $\Omega_{b,0}=0.7$.  
The ratio of bar length to corotation radius is $R=0.7$.  
Frames are separated by one full initial bar rotation period.  
A smaller ratio of bar length to corotation radius results in weaker 
spiral structure during bar growth.  The R2 ring takes longer to become aligned with the bar, 
and the radii of the rings is larger as the resonances have moved outward. 
\label{mon1ceven.eps} 
}
\end{figure}

\begin{figure}
\includegraphics[angle=0,width=3.3in]{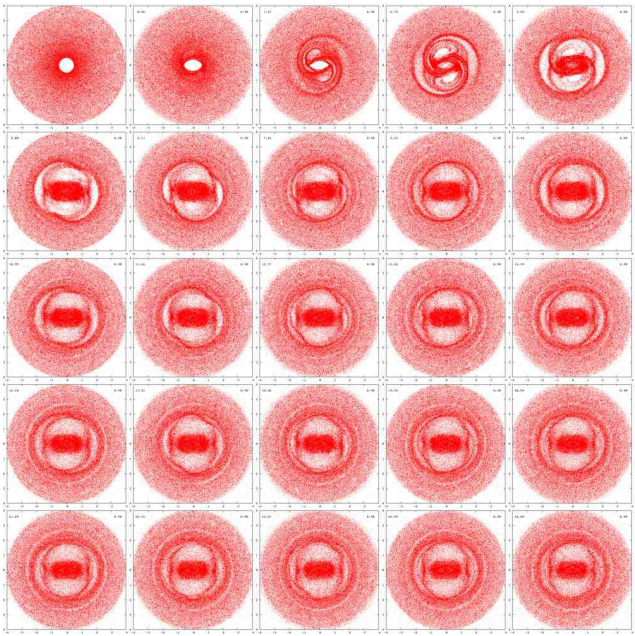}
\caption{
Simulation 36 with an initial pattern speed of $\Omega_{b,0}=0.9$ and $R=0.9$.  
Frames are separated by one full initial bar rotation period.  
We see stronger spiral structure during bar growth, 
and the R1 and R2 rings form and align (perpendicular and parallel, 
respectively) with the bar earlier than in simulations with slower pattern speeds.  
The resonances have moved inward, creating rings with smaller radii.
\label{mon3ceven.eps} 
}
\end{figure}

Finally, we explore the effect of altering the ratio of bar 
length to corotation radius.  \citet{rau08} finds that late 
galaxies have smaller ratios
of bar length to corotation ratios, $R$, and weaker bars.  
\citet{ath92} and \citet{rau08} find that $R=0.7-0.9$ for 
most galaxies.  Previously we have only investigated simulations 
with an initial bar pattern speed of $\Omega_{b,0}=0.8$, 
a corotation radius of $r_{CR}=1.25$, and a ratio of bar length 
to corotation radius of $R=0.8$.  Figure \ref{mon1ceven.eps} shows 
simulation 35 with an initial bar pattern speed of $\Omega_{b,0}=0.7$.  
The ratio of corotation radius to bar length here is $r_{CR}=1.43$ 
and the ratio of bar length to corotation radius is $R=0.7$.  Figure \ref{mon3ceven.eps} shows 
simulation 36 with an initial bar patter speed of $\Omega_{b,0}=0.9$, 
and $R=0.9$.  Both of these figures 
may be compared with simulation 1 shown in Figure \ref{mon2beven.eps}, 
for which $\Omega_{b,0}=0.8$, $r_{CR}=1.25$ and $R=0.8$.  

In Figure \ref{mon1ceven.eps} we note that the spiral-like structure 
at the end of bar growth is not as strong here as it is in 
Figures \ref{mon2beven.eps} and \ref{mon3ceven.eps}.  As was true for  
the simulations with decreasing bar pattern speed following 
bar growth (see Figure \ref{domdtgr2.eps}), transient spiral structure 
during bar growth is weaker at slower pattern speeds.  R1 and R2 rings 
are present 1 to 2 periods later than in Figure \ref{mon2beven.eps} and 
2 to 3 periods later than in Figure \ref{mon3ceven.eps}.  Azimuthal 
variations in density in the rings and shifts in the R2 ring orientation 
are present for up to 6 bar periods following bar growth, whereas they 
are present for only 5 periods in Figure \ref{mon2beven.eps}.  
In Figure \ref{mon3ceven.eps} these variations are only present up 
to the 4$^{th}$ bar period following bar growth, and the R2 ring is 
oriented parallel to the bar at this point.  
The timescale for R2 orientation changes is likely to depend on
the libration timescale in the OLR.  When the pattern
speed is decreased with respect to the bar length, the OLR is further
from the end of the bar and so is likely to have a longer libration timescale
accounting for the increase in the length of
time of R2 ring misalignment seen in simulation 35 (shown in 
Figure \ref{mon1ceven.eps}).

The rings of Figure \ref{mon1ceven.eps} have noticeably larger radii 
than the rings of Figure \ref{mon3ceven.eps}.  The radii of the rings 
shrinks as the corotation radius and the radii of the Lindblad resonances 
is decreased.  Finally, there is a large concentration of particles in the 
L4 and L5 Lagrange points in Figure \ref{mon1ceven.eps}, while 
Figure \ref{mon2beven.eps} shows only a small concentration of particles 
corotating with the bar at the L4 and L5 points.  Figure \ref{mon3ceven.eps} 
shows no such concentrations in the Lagrange points.  As mentioned previously, 
an increase in pattern speed and variations in bar strength
will reduce the number of particles 
that are confined to these points. 

\section{Comparison to ring galaxy morphology}

\subsection{Sample}

Following the lead of \citet{rau08} we compare our simulations to 
B-band galactic images from OSUBSGS.  
We initially considered all galaxies in the OSUBSGS that are classified 
as ring galaxies, however we then restricted our choices of galaxies to those 
that have clearly visible outer rings in the B band images 
and with inclinations below $60^\circ$, so that they could
be compared to the morphology of our simulations.    
We also restricted our comparison galaxies to early type barred spiral 
galaxies as they contain less gas and dust compared to later type
galaxies and so would be more appropriately compared to our collisionless
dissipationless simulations.
We include two additional  galaxies, NGC 4314 and NGC 4548, that are not classified
as ring galaxies, but have strong bars and display R1' type ring morphology.
Our sample of comparison objects consists of 9 spiral galaxies with
morphological classifications ranging from SB0/a to SBbc.
These classifications are based on those put forth by \citet{vau76}.
One of the galaxies is weakly barred, NGC 4457, with maximum $Q_T \sim 0.1$ 
\citep{laurikainen04}.
The galaxies that we compare to our simulations are listed in Table \ref{tab:tab3}
with Hubble type, inclination
with respect to the line of sight, distance, 
H band magnitude from the 2MASS extended source catalog, 
estimated circular velocity, 
bar length, strength and estimated bar rotation periods.  
Bar lengths and strengths ($Q_T$) are taken from the measurements
by \citet{laurikainen04}.
Distances in Mpc from the HyperLeda database \citep{hyperleda03} 
calculated using  velocities
corrected for infall of the Local Group towards Virgo and a Hubble constant
$H_0=70$ km s$^{-1}$ Mpc $^{-1}$.
Inclinations are those from the HyperLeda database \citep{hyperleda03}.
The circular velocity is estimated from the H band magnitude and
the luminosity line width relation by \citet{piercetully92}.


For each galaxy in our comparison sample we ran simulations with bar strengths that 
matched those measured by \citet{laurikainen04}.  
When comparing galaxies to our simulations, we focus on the location, 
orientation and morphology of the spiral arms and outer rings.   
We searched through our bank of simulations for images
that best resembled the outer ring galaxy morphology.
Galaxy images have been corrected for inclination and rotated so that
the bar lies horizontal in our figures.  In some cases
the galaxy images have been flipped so that the galaxy is rotating
counter clockwise and so is
in the same direction as our simulations.

We first compare galaxies with strong R1' pseudoring morphology
to morphology displayed by our simulations during bar growth.
We then focus on galaxies with R1R2 rings.  Finally we explore
simulations that can account for the weak bar and distant
outer ring present in NGC 4457.

\subsection{Comparison of particle simulations to 
galaxies exhibiting similar spiral-like structure}

\begin{figure}
\includegraphics[angle=0,width=3.3in]{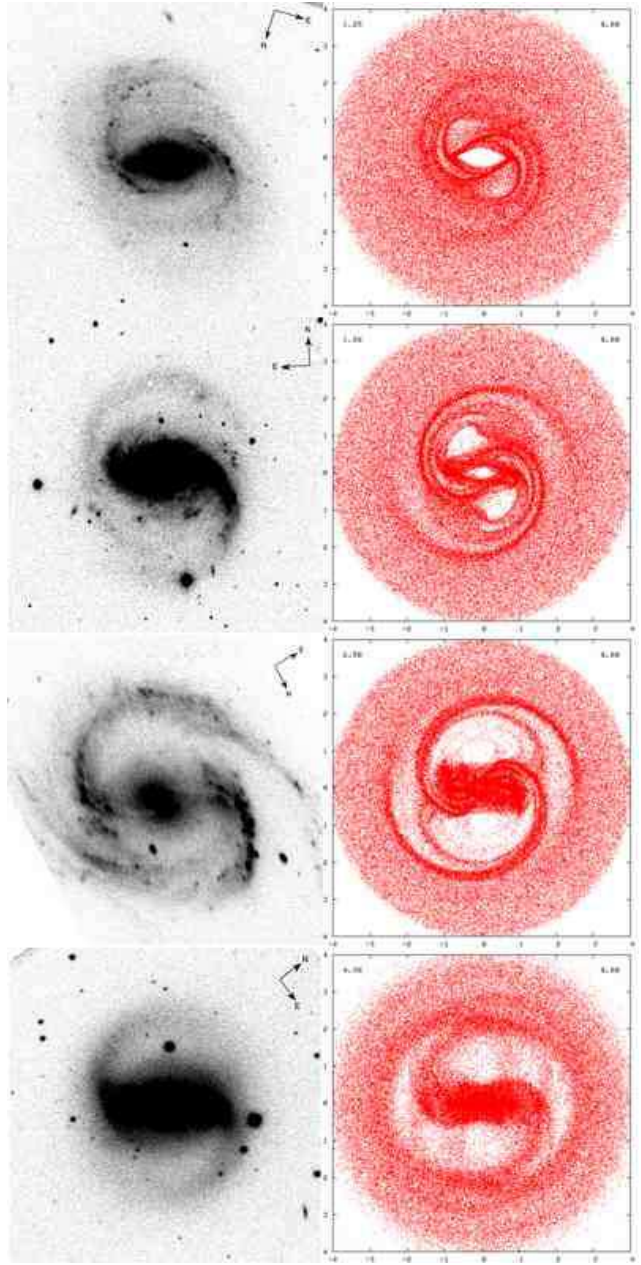}
\caption{
We compare the pseudorings of four galaxies with the structure created during 
bar growth in our simulations.  From top to bottom,
Row 1 shows NGC 4548 with $Q_T=0.34$, Row 2 NGC 7552 with $Q_T=0.4$, Row 3  
NGC 1300 with $Q_T=0.54$, and Row 4 shows NGC 4314 with $Q_T=0.44$.
Galaxy images are on the left and simulations on the right. 
Galaxy images have been corrected for inclination and rotated
so that the bar lies horizontal.  In some galaxies the galaxy image
has been inverted so that the galaxy is viewed rotating counter clockwise.
The simulation for NGC 4548 is shown at $t=1.5P_{b,0}$, (for $P_{b,0}$ initial bar rotation periods) from the beginning of the simulation.  That for 7552 is shown at $t=2P_{b,0}$, that for
NGC 1300 at $t=2.5P_{b,0}$ and that for NGC 4314 at 4$P_{b,0}$.   
The simulation for NGCs 4548, 7552 and 1300
are shown during bar growth, and that for NGC 4314 1 period after bar growth.
The initial orbits for the NGC 4314 simulation had twice the velocity
dispersion of the other simulations.  
\label{b2picsb.eps}
}
\end{figure}

Figure \ref{b2picsb.eps} shows four galaxies, 
with strong R1' type pseudorings; NGCs 4548, 7552, 1300, and 4134.    
To best match their pseudoring morphology, we chose simulations
with $\Omega_{b,0}=0.8$ and a bar that grows in three bar periods.  
Each of the galaxy images is compared with the density 
distribution at a time during or just following bar growth.
NGC 4548 and NGC 7552 have the weakest strengths $Q_T = 0.34$ and 0.4 respectively.
NGC 4314 and NGC 1300 have stronger bars with $Q_T = 0.44$ and 0.54 respectively.
The simulations show stronger spiral structure at late
times during bar growth
and for stronger bars.  Thus we chose a relatively early
time (1.5 bar periods since the beginning of the simulation)
for the simulation matching NGC 4548 which has
weaker spiral structure and later times for the other galaxies 
($t=2P_{b,0}$,  2.5$P_{b,0}$ and 4$P_{b,0}$ for NGC 7552, NGC 1300 and NGC 4314, respectively).
NGC 1300 has the strongest spiral structure 
but also has the strongest bar.   A later time in the simulation 
is required to match the NGC 1300's longer spiral arms. 
NGC 4314's spiral structure is not as narrow as for the other galaxies.
We find a better match of morphologies between 
NGC 4314 and our simulation when we increase the initial velocity dispersion of our simulation from 0.04 to 0.07. The outer disk of NGC 4314 is 
devoid of star formation suggesting that an initial velocity dispersion
typical of a stellar disk rather than a gaseous one should
be used.  The increase in velocity dispersion decreases the strength
of the arms displayed by the simulation, requiring a later
time to match the observed morphology.

We find that our simulations display reasonable matches to observed R1' ring
morphology near the end of bar growth when strong spiral pseudorings are
displayed by our simulations.  
It is likely that these galaxies have experienced recent bar growth. 
These bars could still be growing. 
We note that R1' pseudoring structure is
displayed for a longer timescale (up to 
a few bar rotation periods following bar growth) by SPH and sticky particle
simulations \citep{byrd94,ann00}.  
This comparison suggests that 
our dissipationless simulations could underestimate 
the longevity of R1' structure. 

\subsection{R1R2 rings}

\begin{figure}
\includegraphics[angle=0,width=3.3in]{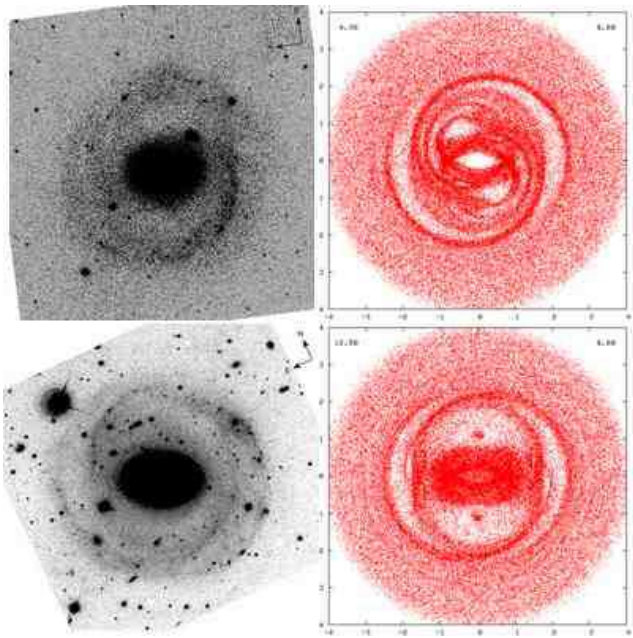}
\caption{
The top row shows NGC 5701, $Q_T=0.14$, compared with the 4$^{th}$ period of
a simulation with a bar of strength $|\epsilon_{tgrow}|=0.07$ 1 period
after bar formation. 
The bottom row shows
NGC 5101, $Q_T=0.19$, compared with the density distribution of a simulation with a bar of strength $|\epsilon_{tgrow}|=0.095$ 7.5 periods
after bar formation. 
The ring structure of these galaxies is similar to the 
broken or misaligned rings shown in our simulations 
a few rotation periods following bar growth. 
\label{b2var.eps}
}
\end{figure}

We now compare our simulations to galaxies that display R1R2 morphology 
which is exhibited by our simulations following bar growth.  We begin by comparing our simulations to two galaxies that exhibit R1R2' morphology.  
Depending on the strength of the bar, the R2 ring can remain misaligned with 
the bar for up to $\sim 10$ bar periods following growth.  
Figure \ref{b2var.eps} shows R1R2' galaxies that exhibit structure similar to that 
in our simulations that is evident before the R2 ring aligns with the bar.  
From top to bottom Figure \ref{b2var.eps} shows NGC 5701, and NGC 5101.

\begin{figure}
\includegraphics[angle=0,width=3.3in]{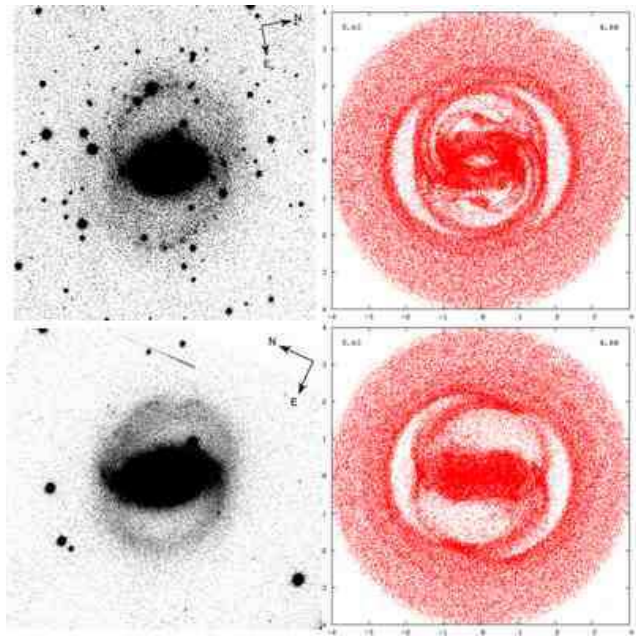}
\caption{
We compare NGC 6782 and NGC 3504 to density distributions of simulations 2 periods after bar growth.
The top row compares NGC 6782, $Q_T=0.17$, to a simulation with a 
bar of strength $|\epsilon_{tgrow}|=0.085$.  
The bottom row shows NGC 3504, $Q_T=0.29$, compared to a simulation 
with $|\epsilon_{tgrow}|=0.145$.  
The bars of these galaxies may be responsible 
for the R1 structure visible in these images.  The galaxies' R2 rings may be 
present but not visible in the B-band.
\label{stbar.eps}
}
\end{figure}

Figure \ref{stbar.eps} compares NGC 6782 and NGC 3504 with density distributions 
showing simulations that have stabilized and exhibit R1R2 morphology. 

NGC 5701 is a fairly weak bar with $Q_T=0.14$ that shows a strong R1 ring
but a weaker R2 ring.  NGC 5701 is compared to a simulation 
one period after bar growth has completed.  At this time
the spiral arms generated during bar growth are beginning to close
and will soon form both R1 and R2 rings. 
As NGC 5701 has a weaker bar it may take longer for R1 and R2 rings to form and become 
aligned perpendicular and parallel with the bar, respectively.  

NGC 5101 is somewhat stronger with $Q_T=0.19$, and displays
both R1 and R2 rings.  It is compared to a simulation
7.5 periods following bar growth. 
The R2 ring at this time is misaligned with the bar similar to
the misalignment in the galaxy.  The galaxy displays a more elliptical
ring than that of our simulation.   A number of factors could account for this
discrepancy.  The bar could be slowing down and causing
increased epicyclic motion in the R2 ring due to resonance capture.  
We may not have corrected for inclination correctly, or the rotation 
curve could be dropping in this region, causing the OLR to be stronger
than we have considered here with a flat rotation curve.

The simulations  we have chosen to match the galaxies with R1R2 morphology
have bars with fixed pattern speeds.  Reasonable matches
between observed and simulated morphology are found a few periods
following bar growth during which time our simulations
contain both R1 and R2 rings but still exhibit asymmetries in
the R2 ring.  Based on
the exploration in section 2.2 we can conclude
that the bars in these galaxies are unlikely to have increased in pattern
speed as this would have destroyed the R1 ring.  Moderate 
decreases in bar pattern speed could have occurred.

We now compare our simulations to two galaxies that exhibit R1R2 morphology.  In Figure \ref{stbar.eps} NGC 6782 and NGC 3504 are both compared 
to the eleventh frame of simulations with bars of strengths 
$|\epsilon_{tgrow}|=0.085$ and 0.145, respectively.  
Both simulations are shown 2 periods after bar growth.  

NGC 6782 is an (R')SB(r)0/a galaxy with an inclination of 
56.0$^\circ$ and $Q_T=0.17$.  
Because of the weaker bar, the R2 ring in NGC 6782's comparison  simulation remains misaligned with the bar for $\sim 6$ bar periods following bar growth.  The spiral-like structure evident during bar growth does not close to form R1 and R2 rings until 2 periods after bar growth, in the frame shown in Figure \ref{stbar.eps}.  There is a reasonable match between the R1 structure of NGC 6782 and that exhibited by our simulation.

NGC 3504 is classified as an (R)SAB(s)ab galaxy with an inclination of 53.4$^\circ$ and $Q_T=0.29$.  NGC 3504 has a much stronger bar than NGC 6782.  R1 and R2 rings are evident only 1 period after bar growth, and the R2 ring is aligned with the bar within 4 bar rotation periods following bar growth.  As with NGC 4314 in Figure \ref{b2var.eps}, the strong bar leads to a weakening of the R1 ring. 

Sticky particles simulations by \citet{schwarz84} displayed
R1 or R2 morphology depending on the initial gas distribution.
Either dissipation is required to exhibit R1 morphology only, 
or the initial stellar and gas distribution is different
for the galaxies showing longer lived R1 morphology.

\subsection{Dissolving the R1 ring and NGC 4457}

We found in section 2.2  that the R1 ring dissolves when the bar pattern speed increases by more than $\sim 8\%$ after bar growth.  We see the same loss of the R1 ring if the bar has strength $|\epsilon_{tgrow}| > 1.5$ or if the bar strength increases by $\gtrsim 140\%$ after bar growth.  In either case, the R1 ring is destroyed leaving a nearly circular R2 ring.  

\begin{figure}
\includegraphics[angle=0,width=3.3in]{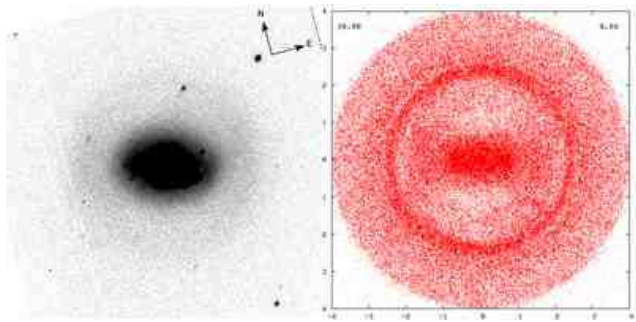}
\caption{
NGC 4457 compared to the 25$^{th}$ frame of simulation 5, which has a bar with increasing pattern speed; $d\Omega_b/dt=0.0004$.  
Both the galaxy and the simulation lack an R1 outer ring 
but do maintain  a nearly circular outer ring.
\label{mon4457.eps}
}
\end{figure}

NGC 4457, an (r)SAB(s)0/a galaxy, 
appears to be lacking an R1 ring.  
According to \citet{laurikainen04} NGC 4457's bar has $Q_T=0.09$, 
which corresponds to a very weak bar.  It is therefore 
unlikely that loss of the R1 ring is due to the strength of the bar, 
as we find that strong bars dissolve the R1 ring and weakening bars
can leave behind a double ring. 
It is possible, however, that the lack of an R1 ring 
is the result of a bar that has increased in pattern speed.
For this reason, in Figure \ref{mon4457.eps} we compare NGC 4457 with the 
last frame ($t=25P_{b,0}$) of simulation 5, which has a bar that is speeding up after bar growth, 
$d\Omega_b/dt=0.0004$.  NGC 4457's outer rings are very faint in the B-band image, 
yet it does appear that the outer ring is circular and no R1 ring is evident.  
It is possible that the pattern speed of this galaxy's 
bar has 
increased since the bar finished growing, thereby destroying any R1 ring that would 
have formed shortly after bar growth.  

\section{Discussion and Summary}

We have presented integrations of collisionless  massless
particles perturbed by growing and 
secularly evolving bar perturbations.  
We find that collisionless simulations can exhibit double
ringed R1 and R2 outer ring
morphology with rings both perpendicular (R1) and parallel (R2) to the bar.  
In the last period of bar growth, strong open spiral structure
is exhibited resembling an R1' pseudoring.   For 2-3 periods
following bar growth R1 and R2 rings are seen with the R2 ring changing
in orientation and azimuthal density contrast.  
Thus R1R2' pseudoring morphology 
is displayed within a few bar periods following bar growth.  
Our simulations start with particles in nearly circular orbits 
with velocity dispersions equivalent to 7 km/s for a 200 km/s rotation curve.
This suggests sticky particle simulations have been successful
in exhibiting R1R2 ring morphology because the velocity
dispersion of orbits is damped and so particles are in initially
nearly circular orbits.

In our collisionless simulations we find that the 
outer rings with major axis perpendicular (R1) 
to the bar are fragile.  
If the bar pattern speed increases more than 8\% after bar growth, 
or if the bar strength is higher than or increases
past $|\epsilon| \gtrsim 0.16$ or $Q_T \gtrsim 0.32$ 
the R1 outer ring will dissolve after $\sim 20$ twenty bar periods.
The simulations are then nearly mirror symmetric and do not display 
asymmetries typical of pseudorings.

Stronger bars can form R1' pseudorings earlier. However if 
the bar strength $|\epsilon| \gtrsim 0.16$ or $Q_T \gtrsim 0.32$ 
the R1 ring will dissolve after $\sim 20$ bar rotation periods.
If the bar strength increases to this value
subsequent to formation, the R1 ring also dissolves. 

We find that a decrease in the bar pattern speed after bar growth causes
particles to be captured in orbits 
parallel to the bar which are increased in epicyclic amplitude
as the bar slows down.  
Strong R1 and elongated R2 rings persist in these simulations.
Misalignments between the R2 ring and the bar also persist so
the galaxy can exhibit R1R2' pseudoring morphology for a longer period
of time.  If the bar pattern speed slows down more than $\sim 3.5\%$
the R2 ring develops a scallop above and below the bar.
As these are not observed in galaxies, bars probably do not
slow down more than $\sim 3.5\%$ without also varying in strength.
  
\citet{sandage94} find that early type barred galaxies often
have semi-detached outer rings (e.g, NGC 1543, \citealt{buta96}
and NGC 4457).
These galaxies may contain bars that have increased
in pattern speed or were once strong and so destroyed their R1 ring. 
If the bar weakens the R1 and R2 rings can be left behind
as two nearly circular rings, similar to those observed in
the unusual double outer ringed galaxy NGC 2273.

We find that the morphology of our simulations resembles
that of R1' ringed galaxies if the simulation time is chosen during or just after
bar formation.  We find we can match
pseudoring morphology with simulations that have 
bar strengths estimated from the bar shapes.  
Stronger and longer spiral arms are seen later in
the simulation and in more strongly barred systems.
The constraint on simulation timescale 
suggests that R1' ring morphology is a signpost of recent bar formation.  
We note that sticky particle and SPH simulations  exhibit
R1 pseudoring morphology a few bar rotation periods
longer than ours suggesting that the dissipationless simulations
explored here underestimate the longevity of these features.

We find that galaxies with R1R2' morphology are well matched by
simulations a few bar rotation periods following bar growth.
As R1 rings are fragile, we infer that these galaxies
have had stable bars that have not experienced large changes
in either pattern speed or strength.

The exploration of parameter space in the collisionless 
dissipationless limit
done here can be used by future work to differentiate between phenomena
that would be exhibited by collisionless models and that that
is a result of dissipation.
A better understanding of the role of dissipation in affecting outer ring morphology
should allow observationally based
constraints on the secular evolution of bars.

Only 10-20\% of early type galaxies exhibit outer rings 
with pseudorings being more prevalent in later type galaxies \citep{buta96}. 
Not all but most galaxies classified with outer rings are barred suggesting
that only 15-40\% of barred galaxies exhibit outer rings.  
Here we have found that R1' and R1R2' galaxies are likely to represent
different times since bar formation with R1' galaxies representing an
earlier timescale during or just after bar formation and R1R2 morphology
representing galaxies with stable bars a few bar rotation periods
following bar formation.  Galaxies in these two transient categories probably
comprise a significant fraction of all outer ring galaxies.  
This suggests that most outer ring galaxies represent morphology that is
only present for a few bar rotation periods.   
It is interesting to ask what timescales these morphologies correspond to.
Bar rotation periods for the ringed galaxies in our sample range from
$\sim 100-200$ Myr (see \ref{tab:tab3}). 
The R1' classification, may only last
a few bar rotations or 1/2 Gyr and the R1R2' classification only $\sim$ 1 Gyr.
Both of these timescales are short compared to the lifetime of a galaxy.
Ringed galaxies lacking R1 rings may be longer lived but may provide
evidence for bar evolution.
It is likely that only a low fraction of barred galaxies
might be considered systems that are not evolving secularly or have not 
formed in the last Gyr. 

\section{Acknowledgments}

We thank Eija Laurikainen for helpful correspondence.
This work made use of data from the Ohio State University 
Bright Spiral Galaxy Survey, which was funded by grants 
AST-9217716 and AST-9617006 from the United States National Science 
Foundation, with additional support from the Ohio State University.
This project was supported in part by NSF award PHY-0552695.
We acknowledge the usage of the HyperLeda database (http://leda.univ-lyon1.fr).
This publication makes use of data products from the Two Micron All Sky Survey, which is a joint project of the University of Massachusetts and the Infrared Processing and Analysis Center/California Institute of Technology, funded by the National Aeronautics and Space Administration and the National Science Foundation.

{}

\begin{table*}
\begin{minipage}{120mm}
\caption{Common Parameters for Simulations} 
\label{tab:tab1}
\begin{tabular}{@{}lccc}
\hline
  & Parameter      & Value & Comments \\
\hline
1 & $\gamma$       & 0.0   & Sets the slope of the rotation curve \\
2 & $t_{grow} $    & 3     & Bar growth time in bar rotation periods \\
3 & $\Omega_{b,0}$ & 0.8   & Initial bar pattern speed \\
4 & $r_{b,0}$      & 1.0   & Initial bar length \\
5 & $r_{CR}$       & 1.25  & Radius of corotation \\
6 & $R$            & 0.8   & Ratio of bar length to corotation radius \\
7 & $\sigma$       & 0.036 & Initial velocity dispersion in units of the circular velocity \\
\hline
\end{tabular}
{ \\
With the exception of simulations 35 and 36, 
these parameters are not altered from simulation to simulation.  
Length scales are given in units of the initial bar length.  
Angular rotation rates are given in units of that at $r_{b,0}$.
The initial bar rotation period is $P_{b,0}=2\pi/\Omega_{b,0}$.  
For simulation 35, $\Omega_{b,0}=0.7$, $r_{CR}=1.43$ and $R=0.7$.  
For simulation 36, $\Omega_{b,0}=0.9$, $r_{CR}=1.1$ and $R=0.9$.
}
\end{minipage}
\end{table*}

\begin{table*}
\begin{minipage}{120mm}
\caption{Additional parameters for Simulations}  
\label{tab:tab2}
\begin{tabular}{@{}lcccccccc}
\hline
& Sim. & $d\Omega_b/dt$ & \%      & $d\Omega_g/dt$ & \%     & $|\epsilon_{tgrow}|$ & $d|\epsilon|/dt$ & \% \\
& 1    & 2              & 3       & 4              & 5      & 6                    & 7              & 8  \\
\hline
& 1    & 0.00           & 0.00    & 0.0            & 0.0    & 0.10                 & 0.0            & 0.0    \\
& 2    & 0.0001         & 2.16    & 0.0            & 0.0    & 0.10                 & 0.0            & 0.0    \\
& 3    & 0.0002         & 4.32    & 0.0            & 0.0    & 0.10                 & 0.0            & 0.0    \\
& 4    & 0.0003         & 6.48    & 0.0            & 0.0    & 0.10                 & 0.0            & 0.0    \\
& 5    & 0.0004         & 8.64    & 0.0            & 0.0    & 0.10                 & 0.0            & 0.0    \\
& 6    & 0.0005         & 10.80   & 0.0            & 0.0    & 0.10                 & 0.0            & 0.0    \\
& 7    & -0.0001        & -2.16   & 0.0            & 0.0    & 0.10                 & 0.0            & 0.0    \\
& 8    & -0.00015       & -3.24   & 0.0            & 0.0    & 0.10                 & 0.0            & 0.0    \\
& 9    & -0.0002        & -4.32   & 0.0            & 0.0    & 0.10                 & 0.0            & 0.0    \\
& 10   & -0.00025       & -5.40   & 0.0            & 0.0    & 0.10                 & 0.0            & 0.0    \\
& 11   & -0.0003        & -6.48   & 0.0            & 0.0    & 0.10                 & 0.0            & 0.0    \\
& 12   & 0.0            & 0.0     & 0.005          & 14.73  & 0.10                 & 0.0            & 0.0    \\
& 13   & 0.0            & 0.0     & 0.01           & 29.45  & 0.10                 & 0.0            & 0.0    \\
& 14   & 0.0            & 0.0     & 0.015          & 44.18  & 0.10                 & 0.0            & 0.0    \\
& 15   & 0.0            & 0.0     & 0.02           & 58.90  & 0.10                 & 0.0            & 0.0    \\
& 16   & 0.0            & 0.0     & -0.0025        & -7.36  & 0.10                 & 0.0            & 0.0    \\
& 17   & 0.0            & 0.0     & -0.005         & -14.73 & 0.10                 & 0.0            & 0.0    \\
& 18   & 0.0            & 0.0     & -0.0075        & -22.09 & 0.10                 & 0.0            & 0.0    \\
& 19   & 0.0            & 0.0     & -0.01          & -29.45 & 0.10                 & 0.0            & 0.0    \\
& 20   & 0.0            & 0.0     & 0.0            & 0.0    & 0.06                 & 0.0            & 0.0    \\
& 21   & 0.0            & 0.0     & 0.0            & 0.0    & 0.08                 & 0.0            & 0.0    \\
& 22   & 0.0            & 0.0     & 0.0            & 0.0    & 0.12                 & 0.0            & 0.0    \\
& 23   & 0.0            & 0.0     & 0.0            & 0.0    & 0.14                 & 0.0            & 0.0    \\
& 24   & 0.0            & 0.0     & 0.0            & 0.0    & 0.16                 & 0.0            & 0.0    \\
& 25   & 0.0            & 0.0     & 0.0            & 0.0    & 0.10                 &  0.0002        & 34.56  \\
& 26   & 0.0            & 0.0     & 0.0            & 0.0    & 0.10                 &  0.0004        & 69.12  \\
& 27   & 0.0            & 0.0     & 0.0            & 0.0    & 0.10                 &  0.0006        & 103.7  \\
& 28   & 0.0            & 0.0     & 0.0            & 0.0    & 0.10                 &  0.0008        & 138.2  \\
& 29   & 0.0            & 0.0     & 0.0            & 0.0    & 0.10                 &  0.0010        & 172.8  \\
& 30   & 0.0            & 0.0     & 0.0            & 0.0    & 0.10                 & -0.0001        & -17.28 \\
& 31   & 0.0            & 0.0     & 0.0            & 0.0    & 0.10                 & -0.0002        & -34.56 \\
& 32   & 0.0            & 0.0     & 0.0            & 0.0    & 0.10                 & -0.0003        & -51.84 \\
& 33   & 0.0            & 0.0     & 0.0            & 0.0    & 0.10                 & -0.0004        & -69.12 \\
& 34   & 0.0            & 0.0     & 0.0            & 0.0    & 0.10                 & -0.0005        & -86.39 \\
& 35   & 0.0            & 0.0     & 0.0            & 0.0    & 0.10                 & 0.0            & 0.0    \\
& 36   & 0.0            & 0.0     & 0.0            & 0.0    & 0.10                 & 0.0            & 0.0    \\
\hline
\end{tabular}
{ \\ 
Length-scales are given in units of the initial bar length.  
Velocities are given in units of the circular velocity at the initial bar end.  
Time is given in units such that the period of rotation is $2 \pi$
at $r_{b,0}$.
Angular rotation rates are in units of that at $r_{b,0}$.  The initial bar period $P_{b,0}=2\pi/\Omega_{b,0}$.
Bar strength is given in units of the square of the circular velocity at $r_{b,0}$.
By Column
Col. (1): Simulation.
Col. (2): $d\Omega_b/dt$ is the rate of bar pattern speed change after bar growth.
Col. (3): The percent by which the bar pattern speed has changed by $t=25 P_{b,0}$, after
twenty-five bar rotation periods.
Col. (4): $d\Omega_g/dt$ is the rate of bar pattern speed change during bar growth.
Col. (5): The percent by which the bar pattern speed has changed by the end of bar growth. 
Col. (6): $\epsilon_{tgrow}$ is the bar strength at the end of bar growth or at time $t=t_{grow}$.
Col. (7): $d\epsilon/dt$ is the bar strength rate of change after $t_{grow}$.
Col. (8): The percent by which the bar strength changes by $t=25 P_{b,0}$. 
Note: Simulations 35 and 36 have initial bar pattern speeds of $\Omega_{b,0}=0.7$ and 0.9, respectively, whereas all other simulations have $\Omega_{b,0}=0.8$ (see Table \ref{tab:tab1}).
}
\end{minipage}
\end{table*}

\begin{table*}
\begin{minipage}{120mm}
\caption{Ring Galaxy Estimated Bar Rotation Periods}
\label{tab:tab3}
\begin{tabular}{@{}lccccccccc}
\hline
         & Type          & $i$   & $Q_T$ & $D$    & $m_H$ & $M_H$ & $v_c$ & $r_b$ & $P_b$ \\
         &               &       &       & Mpc    & mag   &  mag  & km/s  & kpc   & Myr   \\
         &  1            &   2   &   3   &  4     &  5    &  6    & 7     & 8     & 9     \\
\hline
NGC 1300 & (R')SB(s)bc   & 49.3  & 0.54  & 20.1   & 7.770 & -23.7 & 258.6 & 8.5   & 206.7 \\
NGC 3504 & (R)SAB(s)ab   & 53.4  & 0.29  & 23.9   & 8.609 & -23.3 & 234.7 & 7.0   & 187.6 \\
NGC 4314 & SB(rs)a       & 16.2  & 0.44  & 16.4   & 7.725 & -23.3 & 234.7 & 6.0   & 160.7 \\
NGC 4457 & (R)SAB(s)0/a  & 34.6  & 0.09  & 13.4   & 8.015 & -22.6 & 198.1 & 2.9   & 92.0  \\
NGC 4548 & SB(rs)b       & 37.0  & 0.34  & 8.5    & 7.373 & -22.3 & 184.2 & 2.8   & 95.5  \\
NGC 5101 & (R)SB(rs)0/a  & 23.2  & 0.19  & 25.2   & 7.401 & -24.6 & 321.7 & 8.5   & 166.2 \\
NGC 5701 & (R)SB(rs)0/a  & 41.3  & 0.14  & 22.9   & 8.358 & -23.4 & 240.5 & 5.5   & 143.8 \\
NGC 6782 & (R')SB(r)0/a  & 56.0  & 0.17  & 52.0   & 9.115 & -24.5 & 314.0 & 11.7  & 234.4 \\
NGC 7552 & (R')SB(s)ab   & 23.6  & 0.40  & 20.1   & 7.840 & -23.7 & 258.6 & 6.8   & 165.3 \\
\hline
\end{tabular}
{ \\ By Column
Col. (1): Morphological classifications by  \citet{vau76}, except those for 
NGC 1300 and NGC 6782, which are by  \citet{vau91}.
Col. (2): Inclinations from the HyperLeda database \citep{hyperleda03}.
Col. (3): Maximum gravitational bar torque per unit mass per unit square of the circular speed measured by \citet{laurikainen04}.
Col. (4): Distances in Mpc from the HyperLeda database \citep{hyperleda03} calculated using  velocities 
corrected for infall of the Local Group towards Virgo and a Hubble constant 
$H_0=70$ km s$^{-1}$ Mpc $^{-1}$.   
Col. (5): Total integrated flux magnitudes in H band from the
2MASS extended source catalog \citep{jarrett00}. 
Col. (6): Absolute magnitudes in the H band.
Col. (7): Circular velocities estimated from the H-band
magnitude using the luminosity line widths relation 
by \citet{piercetully92}.
Col. (8): Bar lengths measured by \citet{laurikainen04}.  
In their paper \citet{laurikainen04} define the bar length 
to be the radius of the bar region at which the phases of the 
$m=2$ and $m=4$ density amplitudes are constant.
Col. (9): Periods of bar rotation calculated using the angular 
frequencies defined by $\Omega=v_c/r_b$.
}
\end{minipage}
\end{table*}


\begin{thebibliography}{}



\bibitem[Ann \& Lee(2000)]{ann00}
Ann, H. B., \& Lee, H. M. 2000, JKAS, 33, 1	

\bibitem[Athanassoula(1992)]{ath92}
Athanassoula E. 1992, MNRAS, 259, 345

\bibitem[Athanassoula(2003)]{ath03}
Athanassoula, E. 2003, MNRAS, 341, 1179

\bibitem[Bissantz et al.(2003)]{bissantz03}
Bissantz, N., Englmaier, P., \& Gerhard, O.  2003, MNRAS, 340, 949	

\bibitem[Bournaud \& Combes(2002)]{bournaud02}
Bournaud, F., \& Combes, F. 2002, A\&A, 392, 83	

\bibitem[Buta \& Combes(1996)]{buta96}
Buta, R., \& Combes, F.  1996, Fund. Cosmic Physics, 17, 95

\bibitem[Byrd et al.(1994)]{byrd94}
Byrd, G., Rautiainen, P., Salo, H., Buta, R., \& Crocker, D. A.	
1994, AJ, 108, 476	


\bibitem[Combes \& Sanders(1981)]{combes81}
Combes, F., \& Sanders, R. H. 1981, A\&A, 96, 164


\bibitem[Combes \& Gerin(1985)]{combes85}
Combes, F., \& Gerin, M. 1985, A\&A, 150, 327

\bibitem[Contopoulos \& Grosbol(1989)]{cont89}
Contopoulos, G. \& Grosbol, P. 1989, Astr. Astrophys. Rev., 1, 261.

\bibitem[Contopoulos \& Patsis(2006)]{cont06}
Contopoulos, G. \& Patsis, P. A. 2006, MNRAS< 369, 1054 

\bibitem[Das et al.(2003)]{das03}
Das, M., Teuben, P. J., Vogel, S. N., Regan, M. W., Sheth, K., 
Harris, A. I., \& Jefferys, W. H. 2003, ApJ, 582, 190  

\bibitem[Debattista \&  Sellwood(1998)]{debattista98}
Debattista, V. P., \& Sellwood, J. A. 1998, ApJ, 493, L5	


\bibitem[Dehnen(2000)]{dehnen00}
Dehnen, W. 2000,  AJ, 119, 800

\bibitem[de Vaucouleurs et al.(1976)]{vau76}
de Vaucouleurs G., de Vaucouleurs A., \& Corwin, H. G., 1976, 
Second Reference Catalogue of Bright Galaxies,  
University of Texas Press, Austin London

\bibitem[de Vaucouleurs et al.(1991)]{vau91}
de Vaucouleurs G., de Vaucouleurs A., Corwin H. G., Jr, Buta R., Paturel G., Fouque P., 1991, Third Reference Catalogue of Bright Galaxies.  Springer-Verlag, New York

\bibitem[Eskridge et al.(2000)]{eskridge02}
Eskridge, P. B., Frogel, J. A., Pogge, R. W., Quillen, A. C., 
Berlind, A. A., Davies, R. L., DePoy, D. L., Gilbert, K. M., 
Houdashelt, M. L., Kuchinski, L. E., Ram\'irez S. V., Sellgren, K., 
Stutx, A., Terndrup, D. M., Tiede, G. P., 2002, ApJS, 143, 73

\bibitem[Gadotti \& de Souza(2006)]{gadotti06}
Gadotti, D. A., \& de Souza, R. E. 2006, ApJS, 163, 270

\bibitem[Hunter et al.(1988)]{hunter88}
Hunter, J. H., Jr., England, M. N., Gottesman, S. T.,
Ball, R., \& Huntley, J. M. 1988, ApJ, 324, 721

\bibitem[Jarrett et al.(2000)]{jarrett00}
Jarrett, T. H., Chester, T., Cutri, R., Schneider, S., Skrutskie, M., 
Huchra,  J. P. 2000, AJ, 119, 298

\bibitem[Kalnajs(1991)]{kalnajs91}
Kalnajs, A. J. 1991, in Dynamics of Disc Galaxies, 
ed. B. Sundelis,  G\"oteborg, Sweden, p. 323

\bibitem[Laurikainen et al.(2004)]{laurikainen04}
Laurikainen, E., Salo, H., Buta, R., \& Vasylyev, S. 2004, MNRAS,
355, 1251

\bibitem[Lindblad et al.(1996)]{lindblad96}
Lindblad, P. A. B., Lindblad, P. O., \& Athanassoula, E.	
1996, A\&A, 313, 65	

\bibitem[Martinez-Valpuesta et al.(2006)]{martinez06}
Martinez-Valpuesta, I., Shlosman, I., \& Heller, C.
 2006, ApJ, 637, 214


\bibitem[Minchev et al.(2007)]{minchev07}
Minchev, I., Nordhaus, J., \& Quillen, A. C. 2007, ApJ, 664, L31


\bibitem[\protect\citeauthoryear{Paturel et al.}{2003}]{hyperleda03}
Paturel, G., Petit, C., Prugniel, P., Theureau, G., Rousseau, J., 
Brouty, M., Dubois, P., \& Cambr{\'e}sy, L.  2003, A\&A, 412, 45

\bibitem[Pierce \& Tully(1992)]{piercetully92}
Pierce, M. J. \& Tully, R. B. 1992, ApJ, 387, 47	

\bibitem[Quillen(2003)]{quillen03}
Quillen, A. C. 2003, AJ, 125, 785 

\bibitem[Quillen(2006)]{quillen06}
Quillen, A C.,  2006, MNRAS, 365, 1367	

\bibitem[Rautiainen et al.(2008)]{rau08}
Rautiainen, P., Salo, H., \& Laurikainen, E. 2008, MNRAS, in press,
arXiv0806.0471

\bibitem[Rautiainen et al.(2004)]{rau04}
Rautiainen, P., Salo, H., \& Buta, R.	2004, MNRAS, 349, 933	

\bibitem[Rautiainen \& Salo(2000)]{rau00} 
Rautiainen, P., \& Salo, H. 2000, A\&A, 362, 465	

\bibitem[Romero-Gomez et al.(2006)]{romero06}
Romero-Gomez, M.,  Masdemont, J. J., Athanassoula, E., \& Garcia-Gomez, C.	
2006, AA, 453, 39 

\bibitem[Salo et al.(1999)]{salo99}
Salo, H., Rautiainen, P., Buta, R., Purcell, G. B., Cobb, M. L., 
Crocker, D. A., \& Laurikainen, E. 1999, AJ, 117, 792	

\bibitem[Sandage \& Bedke(1994)]{sandage94}
Sandage, A. and Bedke, J. 1994, 
The Carnegie Atlas of Galaxies, Carnegie Inst. of Wash. Publ. No. 638 

\bibitem[Schwarz(1981)]{schwarz81}
Schwarz, M. P. 1981, ApJ, 247, 77

\bibitem[Schwarz(1984)]{schwarz84}
Schwarz, M. P. 1984, Proc. Astr. Soc. Australia, 5, 464


\bibitem[Sellwood \& Debattista(2006)]{sellwood06}
Sellwood, J. A., \& Debattista, V. P. 2006, ApJ, 639, 868


\bibitem[Voglis et al.(2007)]{voglis07}
Voglis,  N., Harsoula, M., \&  Contopoulos, G. 2007, MNRAS, 381, 757


\end{thebibliography}
\end{document}